\documentclass[
 reprint,
 amsmath,amssymb,
 aps,nofootinbib,preprintnumbers,longbibliography,superscriptaddress
]{revtex4-1}

\usepackage[utf8]{inputenc}
\usepackage{graphicx}
\usepackage{dcolumn}
\usepackage{bm}
\usepackage{hyperref}
\usepackage{lipsum}
\usepackage{xcolor}
\usepackage{calc}
\usepackage{accents}
\usepackage{comment}
\usepackage{float}
\usepackage{diagbox}
\usepackage{ulem}

\newcommand\dr{{\rm{dr}}}
\newcommand\dcdm{{\rm{dcdm}}}

\newcommand{\apjl}{ApJ Letters}

\begin{document}

\preprint{LUPM:22-005, YITP-SB-2022-04}

\title{Constraining decaying dark matter with BOSS data \\ 
and the effective field theory of large-scale structures }
\author{Th\'eo Simon}
\email{Electronic address:  theo.simon@umontpellier.fr}
\affiliation{Laboratoire Univers \& Particules de Montpellier (LUPM), CNRS \& Universit\'e de Montpellier (UMR-5299),Place Eug\`ene Bataillon, F-34095 Montpellier Cedex 05, France}

\author{Guillermo Franco Abellán}
\affiliation{Laboratoire Univers \& Particules de Montpellier (LUPM), CNRS \& Universit\'e de Montpellier (UMR-5299),Place Eug\`ene Bataillon, F-34095 Montpellier Cedex 05, France}

\author{Peizhi Du}
\affiliation{C.N. Yang Institute for Theoretical Physics, Stony Brook University, Stony Brook, NY, 11794, USA}

\author{Vivian Poulin}
\affiliation{Laboratoire Univers \& Particules de Montpellier (LUPM), CNRS \& Universit\'e de Montpellier (UMR-5299),Place Eug\`ene Bataillon, F-34095 Montpellier Cedex 05, France}

\author{Yuhsin Tsai}
\affiliation{Department of Physics, University of Notre Dame, IN 46556, USA}

\begin{abstract}

We update cosmological constraints on two decaying dark matter models in light of BOSS-DR12 data analyzed under the Effective Field Theory of Large-Scale Structures (EFTofLSS) formalism, together with {\it Planck}, Pantheon and other BOSS measurements of the baryonic acoustic oscillation (BAO).
In the first model, a fraction  $f_{\rm dcdm}$ of cold dark matter (CDM) decays into dark radiation (DR) with a lifetime $\tau$. 
In the second model (recently suggested as a potential resolution to the $S_8$ tension), all the CDM decays with a lifetime $\tau$ into DR and a massive warm dark matter (WDM) particle, with a fraction $\varepsilon$ of the CDM rest mass energy transferred to the DR. 
Using numerical codes from the recent literature, we perform the first calculation of the mildly non-linear (matter and galaxy) power spectra with the EFTofLSS for these two models.
In the case of DR products, we obtain the constraints $f_{\rm dcdm}\lesssim0.022$ (95\% C.L.) for lifetimes shorter than the age of the universe, and $\tau/f_{\rm dcdm} \gtrsim 250$ Gyr in the long-lived regime assuming $f_{\rm dcdm}\to1$.
We show that {\it Planck} data  contributes the most to these constraints, with EFTofBOSS providing a marginal improvement over conventional BAO and redshift space distortions ($f\sigma_8$) data.
In the case of DR and WDM decay products, we find that EFTofBOSS data significantly improves the constraints at 68\% C.L. on the CDM lifetime with a $S_8$ prior from KiDS-1000.
We show that, in order to fit EFTofBOSS data while lowering $S_8$ to match KiDS-1000, the best-fit model has a longer lifetime $\tau = 120$ Gyr, with a larger kick velocity $v_{\rm kick}/c \simeq \varepsilon \simeq 1.2\%$, than that without EFTofBOSS ($\tau = 43$ Gyr, $\varepsilon =0.6\%$). 
We anticipate that future surveys will provide exquisite constraints on such models.

\end{abstract}

\maketitle

\section{\label{sec:Intro}Introduction}
The $\Lambda$ cold dark matter ($\Lambda$CDM) model provides outstanding explanation for a wide variety of early universe data, such as Cosmic Microwave Background (CMB) and Big Bang Nucleosynthesis (BBN), as well as late universe observations of Large Scale Structure (LSS) including the
Baryon Acoustic Oscillation (BAO), and uncalibrated luminosity distance to SuperNovae of type Ia (SNIa). Despite this remarkable success, the $\Lambda$CDM model does not teach us about the intrinsic nature of its dark sector, made up of both cold dark matter (CDM) and dark energy (DE). In addition, as the accuracy of cosmological observations has improved, the concordance cosmological model starts showing several experimental discrepancies.  The most famous and important cosmological puzzle, the so-called Hubble tension \cite{Verde:2019ivm}, corresponds to a large discrepancy ($\sim 4-5\sigma$) between the local determination of $H_0$ from a variety of methods -- and in particular the cosmic distance ladder based on cepheid-calibrated SNIa by the SH0ES team \cite{Riess:2021jrx}--, and its determination using CMB data under the assumption that the universe is described by the $\Lambda$CDM model \cite{Aghanim:2018eyx}. Another intriguing cosmological conundrum, the one at the heart of this study, is a less significant but older tension ($\sim 2-3\sigma$) between the weak-lensing\footnote{More precisely, there even exists a `lensing is low' anomaly when comparing galaxy clustering and weak lensing data within the $\Lambda$CDM cosmology \cite{Leauthaud:2016jdb,Lange:2020mnl,Amon:2022ycy}.} \cite{KiDS:2020suj,DES:2021wwk,HSC:2018mrq,Amon:2022ycy} and CMB \cite{Aghanim:2018eyx,ACT:2020gnv} determinations of the amplitude of the local matter fluctuations, parameterized as $S_8 = \sigma_8 \sqrt{\Omega_m/0.3}$, where $\Omega_m$ is the current total matter abundance, and $\sigma_8$ corresponds to the root mean square of matter fluctuations on a $8 \ h^{-1}$Mpc scale, with $h = H_0/(100 \ \textrm{km/s/Mpc})$, and is defined as follows
\begin{equation}
    \sigma_8^2 = \int \frac{k^3}{2\pi^2}P_m(k)W_8^2(k)d\ln{k}.
    \label{sigma8}
\end{equation}
Here $P_m(k)$ is the linear matter power spectrum, and $W_8(k)$ is a window function describing a sphere (in Fourier space) with a (historically chosen) radius of $8 \ h^{-1}$Mpc.

\

Barring unknown systematic errors (see e.g. \cite{Freedman:2021ahq,Riess:2021jrx,Amon:2022ycy} for discussion), these discrepancies might be the first clue about the intrinsic nature of the $\Lambda$CDM dark sector. On the one hand, the resolution of the Hubble tension most likely involves new physics in the pre-recombination era\footnote{We note that recent analysis based on the equality scale $k_{\rm eq}$ seems to disfavor some of the most extreme models suggested to resolve the tension and could eventually provide a challenge to early-universe models \cite{Farren_2022,Philcox_2022}.}, through a decrease of the sound horizon before recombination \cite{Bernal:2016gxb,Aylor:2018drw,Knox:2019rjx,Camarena:2021jlr,Efstathiou:2021ocp,Schoneberg:2021qvd}, such as model involving dark radiation and/or new neutrino properties \cite{Kreisch:2019yzn,Berbig:2020wve,Ghosh:2019tab,Forastieri:2019cuf,Escudero:2019gvw,Escudero:2021rfi,Blinov:2020hmc,Ghosh:2021axu,Archidiacono:2022ich,Aloni:2021eaq}, early dark energy \cite{Karwal:2016vyq,Poulin:2018cxd,Smith:2019ihp,Niedermann:2019olb,Niedermann:2020dwg,Ye:2020btb}, modified gravity \cite{Renk:2017rzu,Umilta:2015cta,Ballardini:2016cvy,Rossi:2019lgt,Braglia:2020iik,Zumalacarregui:2020cjh, Abadi:2020hbr,Ballardini:2020iws,Braglia:2020bym,DiValentino:2015bja,Bahamonde:2021gfp,Raveri:2019mxg,Yan:2019gbw,Frusciante:2019puu,SolaPeracaula:2019zsl,SolaPeracaula:2020vpg,Ballesteros:2020sik,Braglia:2020auw,Desmond:2019ygn,Lin:2018nxe} or exotic recombination \cite{Chiang:2018xpn,Hart:2019dxi,Sekiguchi:2020teg,Jedamzik:2020krr,Cyr-Racine:2021alc} (for review, see Refs.~\cite{DiValentino:2021izs,Schoneberg:2021qvd}). On the other hand, the resolution of the $S_8$ tension requires a suppression in the matter power spectrum for $k \sim 0.1 - 1 \ h\,\text{Mpc}^{-1}$ in order to reduce the value of the $\sigma_8$ parameter (see Eq. \ref{sigma8}), which can be achieved through a number of models that take into account new hypothetical properties of dark matter (DM) and/or DE \cite{Lesgourgues:2015wza,Buen-Abad:2015ova,Chacko:2016kgg,Buen-Abad:2017gxg,Heimersheim:2020aoc,Cyr-Racine:2021alc,Gomez-Valent:2020mqn,DiValentino:2019ffd,Lucca:2021dxo,Abellan:2020pmw,DiValentino:2020vvd,Bansal:2021dfh}. 

\

Decaying cold dark matter (DCDM) models, in which dark matter is unstable on cosmological time-scale and decays into invisible products, have been proposed as potential resolutions to cosmic tensions \cite{Enqvist:2015ara,Berezhiani:2015yta,Blinov:2020uvz,vattis_late_2019,Abellan:2020pmw,abellan_2021}. 
In the past it was found that DM models with purely radiation decay products can neither resolve the Hubble tension nor the $S_8$ tension \cite{Chudaykin:2016yfk,Chudaykin:2017ptd,Poulin_2016,Clark:2020miy,Haridasu:2020xaa,Nygaard:2020sow,Alvi:2022aam}, while DM models with massive decay products can resolve the $S_8$ tension, as the massive particle produced during the decay act as a WDM component, reducing power on scale below the free-streaming length at late-times \cite{Abellan:2020pmw,abellan_2021}.  Beyond recent observational tensions, the study of these models is important from the particle physics point of view, as it addresses the question of the stability of DM on long cosmological time-scales. In the literature, there are many models involving the existence of DM decays at late-times, such as models with R-parity violation \cite{BEREZINSKY1991382,KIM200218}, super Weakly Interacting Massive particles (super WIMPs) \cite{CoviEtAl1999,FengEtAl2003,FengEtAl2003b, AllahverdiEtAl2015}, sterile neutrinos \cite{Abazajian:2012ys,Adhikari:2016bei}, models with an additional U(1) gauge symmetry \cite{Chen:2008yi,Choi:2020tqp,Choi:2020nan,Choi:2021uhy}, or more recently a model of decaying warm dark matter \cite{holm2022decaying}. Besides cosmic tensions, some DCDM models were proposed as a way to explain the excess of events in the electronic recoils reported by the Xenon1T collaboration \cite{Choi:2020udy,Xu:2020qsy,Dutta:2021nsy,Abellan:2020pmw,abellan_2021}. In addition, DCDM models with massive daughters have also been suggested as a potential solution to the small (subgalactic) scales structure problem of CDM (e.g.~\cite{LinEtAl2001,SanchezSalcedo2003,CembranosEtAl2005,Kaplinghat2005,StrigariEtAl2007c,BorzumatiEtAl2008,PeterEtAl2010a,PeterEtAl2010,Choi:2020nan}).

\

In this article, we deal with DCDM with two type of decay products: (i) the DCDM $\to$ DR model, where the decay products is only composed of a (massless) dark radiation (DR) component, and (ii) the DCDM $\to$ WDM+DR model, where the decay products are one massive WDM component and one DR component. Previous works have limited themselves to the impact of DCDM decay at the background and linear perturbations level, deriving constraints (and hints) on these models from a combination of {\it Planck} CMB, BAO and uncalibrated luminosity distance to SN1a data.
Here, we go beyond previous works by making use of the Effective Field Theory of Large Scale Structures (EFTofLSS) to describe the mildly non-linear regime of the galaxy clustering power spectrum and derive improved constraints thanks to the EFTofLSS applied to BOSS data. 
The main objectives of this paper are: i) perform the first-ever computation of the mildly non-linear regime in DCDM models with massive and massless decay products through the EFTofLSS; ii) test whether current BOSS data can lead to stronger constraints on these models; and iii) check whether these constraints can put pressure on DCDM models that resolve the $S_8$ tension. 

\

Our paper is structured as follows: in Sec.~\ref{sec:EFT}, we briefly review the EFTofLSS formalism, the observable at hand and the public codes available to perform our analyses; in Sec.~\ref{sec:NL}, we introduce the models and present the non-linear power spectrum computed with the EFTofLSS; in Sec.~\ref{sec:DataAnalysis}, we present the results of comprehensive monte-carlo markov chain analyses of the DCDM model and discuss the implications of these constraints for the $S_8$ tension; we eventually conclude in Sec.~\ref{sec:conclusion}. App.~\ref{sec:app_EFT_vs_Nbody} is dedicated to comparing results of the EFTofLSS with N-body simulations in the DCDM$\to$DR model, while App.~\ref{sec:app_EFT_WDM} details the scope of our computation in the DCDM$\to$WDM+DR model. Finally App.~\ref{sec:app_S8}, App. \ref{sec:app_chi^2} and App.~\ref{sec:app_WDM_LCDM} present additional results of the MCMC analyses for completeness.

\section{\label{sec:EFT}The galaxy power spectrum from the EFTofLSS formalism}

Although an exhaustive review of the EFTofLSS is beyond the scope of this paper\footnote{The first formulation of the EFTofLSS was carried out in Eulerian space in Refs. \cite{Carrasco_2012,baumann_2012} and in Lagrangian space in \cite{porto_2014}. Once this theoretical framework was established, many efforts were made to improve this theory and make it predictive, such as the understanding of renormalization \cite{Pajer_2013, Abolhasani_2016}, the IR-resummation of the long displacement fields \cite{senatore_2014redshift, Baldauf_2015, Senatore_2015IR, senatore_2018IR, Lewandowski_2020, Blas_2016}, and the computation of the two-loop power spectrum \cite{Carrasco_2014twoloop, Carrasco_2014twoloop2}. Then, this theory was developed in the framework of biased tracers (such as galaxies and halos) in Refs. \cite{senatore_2015, Mirbabayi_2015, Angulo_2015, Fujita_2020, perko_2016, Nadler_2018}.}, in this section, we briefly discuss the software tools that are available in the literature making use of the EFTofLSS to analyze the full-shape of the galaxy clustering power spectrum as measured by BOSS. The relevant observables are the multipoles of the galaxy power spectrum, which are obtained through Legendre polynomials ($\mathcal{L}_l$) decomposition:
\begin{equation}
    P_l(z,k)=\frac{2l+1}{2}\int^1_{-1}d\mu\mathcal{L}_l(\mu)P_{\text{gg}}(z,k,\mu),
\end{equation}
where $z$ is the redshift, $\mu= \hat{z}\cdot\hat{k}$ is the the angle between the line-of-sight $\bm{z}$ and the wavevector of the Fourier mode $\bm{k}$, and $P_{\text{gg}}(z,k,\mu)$ is the redshift-space (non-linear) galaxy power spectrum at one-loop order (see the appendix of Ref. \cite{D_Amico_2021} for the formal expression). 
This expression includes  the ``Alcock-Paczynski transformation'' which takes into account the fact that the observation uses artificial cosmological parameters to convert redshifts as well as celestial coordinates into Cartesian coordinates.
The two main contributions to $P_{\text{gg}}(z,k,\mu)$ are the monopole ($l=0$) and the quadrupole ($l=2$).
Currently, there are two codes in the literature that model non-linear effects on the power spectrum at one-loop (including a proper infrared resummation (IR) \cite{senatore_2014redshift, Baldauf_2015, Senatore_2015IR, senatore_2018IR, Lewandowski_2020, Blas_2016} and a number of observational systematics corrections beyond the Alcock-Paczynski effect \cite{Alcock_1979}, such as window functions \cite{Beutler_2019} and fiber collisions \cite{Hahn_2017}) through the EFTofLSS method and which allows us to determine the monopole and the quadrupole of the galaxy power spectrum: i) the PyBird\footnote{\url{https://github.com/pierrexyz/pybird}} code \cite{D_Amico_2021}-- a python module that determines the non-linear matter power spectrum from the linear one returned by a Boltzmann code such as CLASS\footnote{\url{https://lesgourg.github.io/class_public/class.html}} \cite{Blas_2011} or CAMB\footnote{\url{https://camb.info/}} \cite{Lewis:1999bs}, and ii) the CLASS-PT\footnote{\url{https://github.com/Michalychforever/CLASS-PT}} code \cite{Chudaykin_2020}--  which is a stand-alone extension of the CLASS code. Both codes take into account the same effects with respect to a standard linear Boltzmann code, and in particular make use of the ``FFTLog method'' \cite{Hamilton:1999uv,Simonovic:2017mhp} to compute the one-loop power spectrum and the IR resummation.  Given that our DCDM $\to$ WDM+DR study makes use of an independent extension to the CLASS code, we will rely on the PyBird code. We provide a comparison between the two codes in the context of the DCDM$\to$DR model (already implemented in CLASS-PT) in App.~\ref{sec:app_EFT_vs_Nbody}. One might wonder whether the EFTofLSS formalism must be extended to properly described the models under consideration. In App.~\ref{sec:app_EFT_vs_Nbody} and App.~\ref{sec:app_EFT_WDM}, we argue that the current formalism (and the codes in their standard form) is sufficient to describe the DCDM models given present constraints and precision of the data. Yet, we anticipate that the formalism will need to be developed further for future surveys such as Euclid \cite{Amendola:2016saw} and the LSST/Vera Rubin Observatory (VRO) \cite{Mandelbaum:2018ouv}, which will reach sub-percent precision.

\

The data we use, in order to confront the non-linear galaxy power spectrum forecasts with the observations, are made of three different sky-cut from BOSS DR12 \cite{Alam:2016hwk, reid_2015sdssiii,Kitaura_2016}:  LOWZ NGC, CMASS NGC and CMASS SGC. LOWZ corresponds to the the BOSS DR12 data including the BAO post-reconstruction for $0.2 < z < 0.43$ and has an effective redshift $z_{eff,\text{LOWZ}}=0.32$, while CMASS corresponds to the the BOSS DR12 data also including the BAO post-reconstruction for $0.43 < z < 0.7$ and has an effective redshift $z_{eff,\text{CMASS}}=0.57$ (see Ref. \cite{zhang2021boss}). This data set will be called, in the following, `EFTofBOSS data'. Finally, it is worth noting that the EFTofLSS method has been tested against various simulations (\cite{D_Amico_2021, d_Amico_2020, Colas_2020, Nishimichi_2020}), and it has been highlighted that the BOSS full shape can only be evaluated up to $k_{\text{max}} \sim 0.2 \ h\,\text{Mpc}^{-1}$, where the BOSS full shape corresponds to the combination of the monopoles and quadrupoles of the power spectra of LOWZ NGC, CMASS NGC and CMASS SGC. 
To be more precise, we consider that $k_{\text{max,LOWZ}} = 0.2 \ h\,\text{Mpc}^{-1} $ and $k_{\text{max,CMASS}} = 0.23 \ h\,\text{Mpc}^{-1}$. 
Finally, we mention that the PyBird code makes use of ten additional nuisance parameters per sky-cut to describe various aspects of the EFTofLSS (for more details see e.g.\cite{perko_2016}):
\begin{itemize}
\item  4 parameters $b_i$ ($i=1,2,3,4$) to describe the galaxy bias at one-loop order;
\item 3 parameters $c_{ct}$, $c_{r,1}$, and $c_{r,2}$ corresponding to counterterms. $c_{ct}$ is a linear combination of a higher derivative bias and the dark matter sound speed, while $c_{r,1}$ and $c_{r,2}$ are the redshift-space counterterms;
\item 3 parameters $c_{\epsilon,0}$, $c_{\epsilon,1}$ and $c_{\epsilon,2}$ which describe stochastic contributions.
\end{itemize}
In practice, we make use of the analytical marginalization of Ref.~\cite{DAmico:2020kxu} (app.~C) \footnote{When discussing best-fits however, we also optimize the nuisance parameters that are analytically marginalized in the MCMC.} such that only two extra parameters per sky-cut are required in the analysis.

\section{\label{sec:NL}Non-Linear power spectrum in DCDM cosmologies}

In this section, we review the models of decaying dark matter considered in this work, and present the first computation of the non-linear power spectra in these cosmologies. We consider two different DCDM models (both are limited to decay into the dark sector): one in which a fraction of dark matter decays into massless particles, and the second one in which all of the dark matter experiences two-body decay into massive and massless particles.

\subsection{Dark Radiation decay products (DCDM $\to$ DR model)}

\subsubsection{Presentation of the model}

In the first model we consider, the cold DM sector is  partially composed of an unstable particle (denoted as DCDM) that decays into a non-interacting relativistic particle (denoted as DR). The rest of the DM is considered stable and we refer to it as the standard CDM. 
In addition to the standard six $\Lambda$CDM parameters, there are two free parameters describing the lifetime of DCDM $\tau$ (or equivalently the decay width $\Gamma = \tau^{-1}$), as well as the fraction of DCDM to total dark matter at the initial time $a_{\rm ini}\to 0$:
\begin{equation}
    f_{\rm dcdm} \equiv\frac{\omega_{\rm dcdm}(a_{\rm ini})}{\omega_{\text{tot,dm}}(a_{\rm ini})},
    \label{def_f}
\end{equation} 
with  $\omega_{\text{tot, dm}}\!\equiv\! \omega_{\text{dcdm}} + \omega_{\text{cdm}}$. With these definitions, in the limit of large $\tau$ and/or small $f_{\rm dcdm}$, one recovers the $\Lambda$CDM model.

\

The evolution of the homogeneous energy densities of the  decaying dark matter and dark radiation is given by (see e.g. Refs. \cite{Audren:2014bca,Enqvist_2015,Poulin_2016}):
\begin{align}
    \dot{\bar{\rho}}_{\text{dcdm}} + 3 \mathcal{H}\bar{\rho}_{\text{dcdm}} &= -a\Gamma \bar{\rho}_{\text{dcdm}}, \\
\dot{\bar{\rho}}_{\text{dr}} + 4 \mathcal{H}\bar{\rho}_{\text{dr}} &= a\Gamma \bar{\rho}_{\text{dcdm}},
\end{align}
where $\mathcal{H}$ is the conformal Hubble parameter,
\begin{equation}
    \mathcal{H}^2(a)=\frac{8\pi G a^2}{3} \sum_i\bar{\rho}_i(a),
\end{equation}
with
\begin{align}
    \sum_i\bar{\rho}_i(a) &= \bar{\rho}_{\text{cdm}}(a) + \bar{\rho}_{\text{dcdm}}(a) + \bar{\rho}_{\text{dr}}(a) \nonumber \\
    &+ \bar{\rho}_\gamma(a) + \bar{\rho}_\nu(a) + \bar{\rho}_b(a) + \bar{\rho}_{\Lambda}.
\end{align}

\

To describe the evolution of the linearly perturbed universe, we consider the usual synchronous gauge, where the scalar part of the perturbed metric is written as \cite{Ma_1995}
\begin{equation}
    ds^2=a^2(\tau)\left[-d\tau^2 + (\delta_{ij}+h_{ij}(\bm{x},t))dx^idx^j\right]\,.
\end{equation}
Here $\tau$ is the conformal time, and $h_{ij}(\bm{x},\tau)$ is defined as
\begin{align}
    h_{ij}(\bm{x},\tau) &= \int \ d^3k e^{i\bm{k}.\bm{x}} [ \hat{k}_i \hat{k}_j h(\bm{k},\tau) \nonumber \\
    &+ \left(\hat{k}_i \hat{k}_j - \frac{1}{3}\delta_{ij}\right) 6 \eta(\bm{k},\tau)]\,.
\end{align}
$h$ denotes the trace of $h_{ij}$, while $\eta$ corresponds to the other traceless scalar degree of freedom of the metric perturbation in Fourier space. Additionally, we consider the frame co-moving with the DCDM (and CDM) fluid, such that  $\theta_{\text{dcdm}}=\partial_iv^i_{\text{dcdm}} = 0$, where $\theta_{\text{dcdm}}$ is the divergence of the DCDM velocity $v^i_{\text{dcdm}}$. As a result, the energy density perturbation of the DCDM component, $\delta_{\text{dcdm}} \equiv \rho_{\text{dcdm}}/\bar{\rho}_{\text{dcdm}}-1$, follows the same evolution as standard CDM:
\begin{equation}
    \dot{\delta}_{\text{dcdm}}= -\frac{\dot{h}}{2}.
    \label{delta_DCDM_evolution}
\end{equation}
The evolution of the linear perturbations of the DR integrated phase-space distribution multipoles is governed by the following hierarchy of equations \cite{Audren:2014bca,Enqvist_2015,Poulin_2016}:
\begin{align}
\dot{F}_{\dr,0}   &= - k F_{\dr,1} 
 -\frac{2}{3} r_{\rm dr} \dot{h}+\dot{r}_{\rm dr} \delta_\dcdm, \label{delta_DR_evolution_1} \\ 
\dot{F}_{\dr,1}   &= \frac{ k}{3}F_{\dr,0}-\frac{2k}{3} F_{\dr,2}, \label{theta_DR_evolution_1}  \\ 
 \dot{F}_{\dr,2}  &= \frac{2 k}{5} F_{\dr,1}-\frac{3 k}{5} F_{\dr,3}
 + \frac{4}{15}r_{\rm dr} (\dot{h}+6\dot{\eta}),  \label{shear_DR_evolution_1}  \\ 
 \dot{F}_{\dr,\ell} &= \frac{ k}{(2\ell+1)} \left[ \ell F_{\dr,\ell-1}-(\ell+1) F_{\dr,\ell+1} \right] \ \ \ \ \  \ (\ell \geq 3). \label{l3_DR_evolution_1} 
\end{align}
In the previous equations we have introduced $r_\dr \equiv a^4 \bar{\rho}_\dr(a) /\rho_{c,0}$ following Ref.~\cite{Poulin_2016}, where $\rho_{c,0}$ is the critical density today. In the scenario under study, we have:
\begin{equation}
\dot{r}_\dr = a \Gamma 
(\bar{\rho}_\dcdm/\bar{\rho}_\dr) r_\dr. 
\end{equation}
We also note that the first three multipoles are simply related to elements of the perturbed stress-energy tensor as $F_{\dr,0} = r_\dr \delta_\dr$, $ F_{\dr,1} =(4 r_\dr/3 k )\theta_\dr$, and $ F_{\dr,2} = 2 \sigma_\dr r_\dr$. 
In order to truncate the hierarchy of Eqs.~(\ref{delta_DR_evolution_1})-(\ref{l3_DR_evolution_1}) at some $\ell_{\rm max}=17$, we adopt the scheme proposed in Ref.~\cite{Ma_1995} for massless neutrinos (and extended in CLASS to include non-zero curvature \cite{Lesgourgues:2013bra}) in order to limit the propagation of the error from $\ell_{\rm max}$ to $\ell$. We extrapolate the behavior of  $F_{\rm dr,\ell_{\rm max}+1}$ thanks to the recursion relation:
\begin{equation}
    F_{\rm dr, \ell_{\rm max}+1} \approx \frac{2\ell_{\rm max}+1}{k\tau} F_{\rm dr, \ell_{\rm max}}- F_{\rm dr, \ell_{\rm max}-1}.\end{equation}

\

These equations have been implemented in the Boltzmann code CLASS, and the impact of DCDM $\to$ DR decay on the (linear) CMB and matter power spectrum has been studied in details in the literature \cite{Audren:2014bca,Enqvist_2015,Poulin_2016}. In App. \ref{sec:app_EFT_vs_Nbody} we present a comparison of the EFTofLSS calculation with N-body simulations performed in Ref.~\cite{hubert2021decaying}. The results obtained from these two methods agree upto sub-percent difference for $k\lesssim 0.2 \ h\,\text{Mpc}^{-1}$ and $z=0$, justifying that one can safely analyze the (mildly) non-linear galaxy power spectrum with the EFTofLSS.

\subsubsection{The non-linear power spectrum}
\label{sec:NLDR}

\begin{figure*}
    \centering
    \includegraphics[trim=2cm 0cm 3cm 11cm, clip=true, width=2\columnwidth]{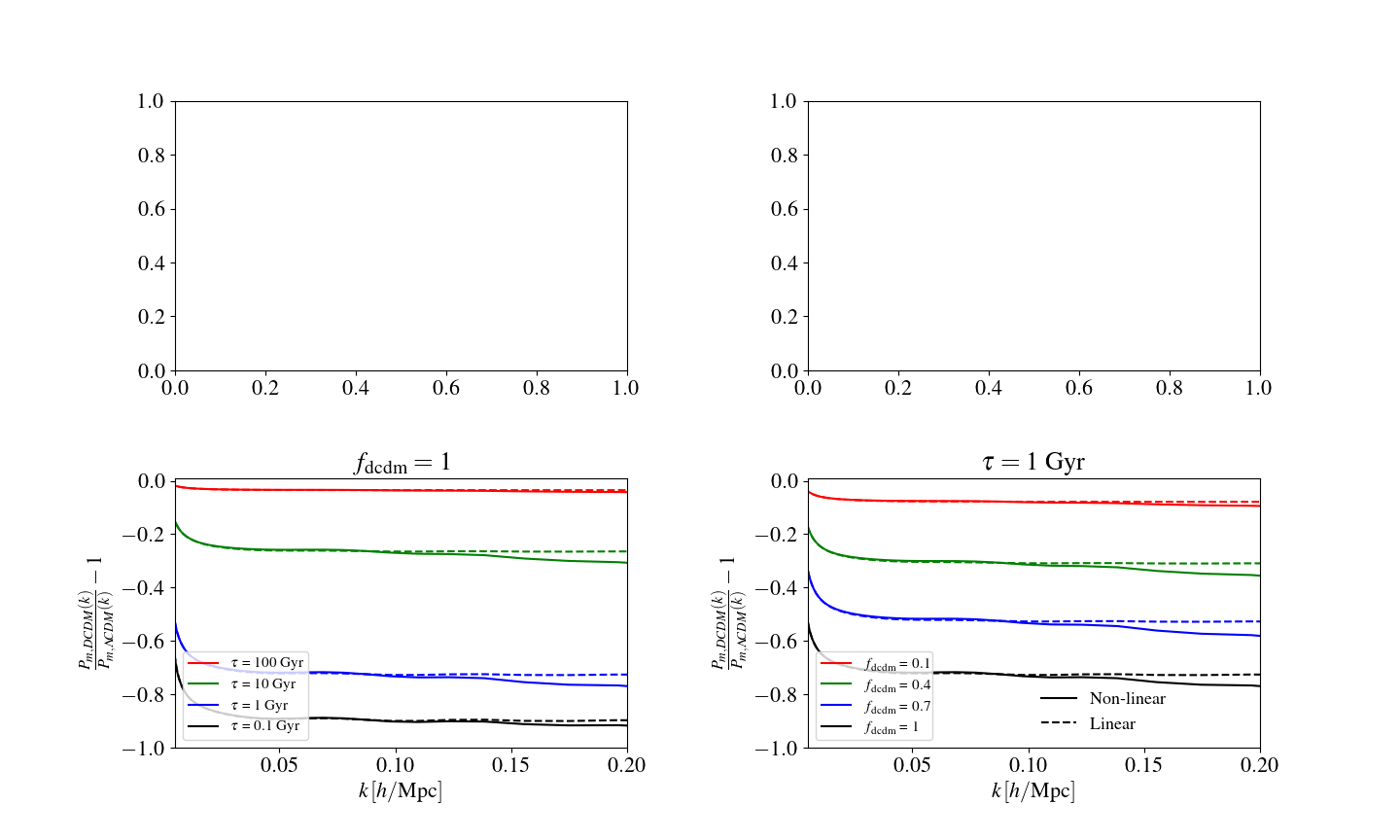}
    \caption{\textit{Left} - Residuals of the linear (dashed lines) and non-linear matter power spectrum (solid lines) for $f_{\rm dcdm}$ set to 1 and $\tau =$ 0.1, 1, 10, 100 Gyr. Residuals are taken with respect to the $\Lambda$CDM model at $z=0$. \textit{Right} - The same, but this time $\tau$ is set to 1 Gyr and $f_{\rm dcdm} = $ 0.1, 0.4, 0.7, 1.}
    \label{fig:DR_linear}
    \centering
    \includegraphics[trim=2cm 0cm 3cm 0cm, clip=true, width=2\columnwidth]{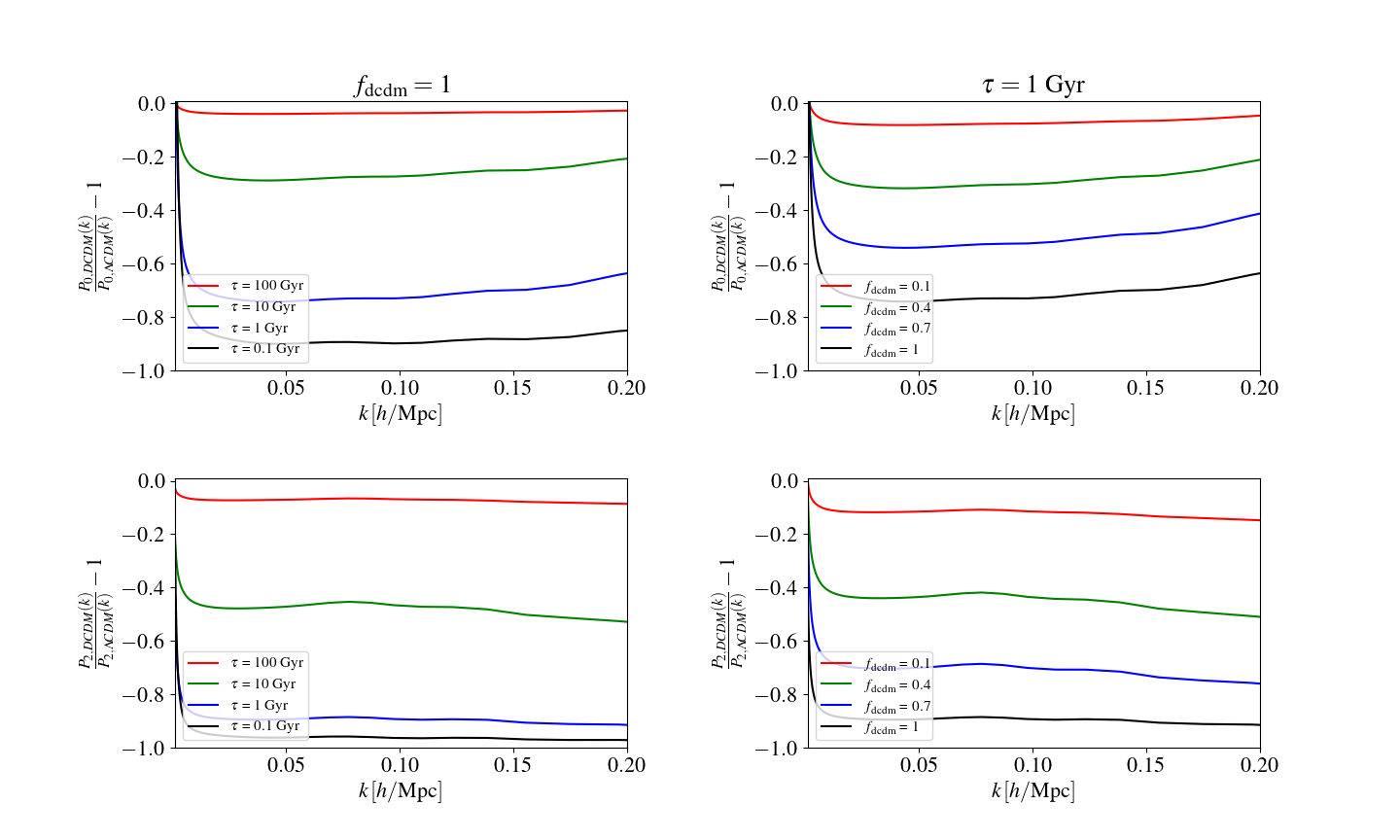}
    \caption{\textit{Left} - Residuals of the monopole and the quadrupole of the galaxy power spectrum for $f_{\rm dcdm}$ set to 1 and $\tau =$ 0.1, 1, 10, 100 Gyr. Residuals are taken with respect to the $\Lambda$CDM model at $z=0$. \textit{Right} - The same, but this time $\tau$ is set to 1 Gyr and $f_{\rm dcdm} = $ 0.1, 0.4, 0.7, 1.}
    \label{fig:pybird_dcdm_dr}
\end{figure*}

Thanks to the PyBird code, we plot in Fig. \ref{fig:DR_linear} the residuals of the non-linear matter power spectra of the DCDM $\to$ DR model with respect to that of the $\Lambda$CDM model at $z=0$. We also represent the associated linear matter power spectra obtained from the CLASS code. In addition, we plot in Fig. \ref{fig:pybird_dcdm_dr} the residuals of the monopole and quadrupole of the galaxy power spectra of this model. In these figures, we set the $\Lambda$CDM parameters\footnote{For completeness, note that the shape of the residuals of the galaxy and matter power spectra depend on the values of the EFT nuisance parameters, especially at large $k$. According to the notation of Ref. \cite{D_Amico_2021}, for the numerical evaluation we set the effective dark matter sound speed $c_{s} = 1$ for the matter power spectra, and $b_1 = 2$, $b_2 = 1$, $b_3 = 0.5$, $b_4=0$, $c_{ct} = 0.5$, $c_{r,1}= 2$ and $c_{r,2}= c_{\epsilon,0} = c_{\epsilon,1} = c_{\epsilon,2} = 0$ for the galaxy power spectra. In practice, these parameters are optimized when quoting best-fits, to ensure that they take realistic values.} to their best-fit values from the analysis of  `{\it Planck} + Pantheon + EFTofBOSS + Ext-BAO' (as described in sec.~\ref{sec:DataAnalysis}). Finally, we simply vary the two parameters $f_{\rm dcdm}$ and $\tau$ to isolate their cosmological effects : in the left panels, we fix $f_{\rm dcdm}=1$ and vary $\tau\in [0.1,1000]$ Gyr, while in the right panel  we fix $\tau = 1$ Gyr and vary $f_{\rm dcdm}\in[0.1,1]$.

\

From Figs.~\ref{fig:DR_linear} and \ref{fig:pybird_dcdm_dr}, one can see that the monopole of the galaxy power spectrum shows a behavior very similar to that of the linear matter spectrum. For a realistic choice of EFT parameters, it shows an almost scale-independent power suppression due to two main reasons \cite{Audren:2014bca,Poulin_2016}. First, the decay of DCDM decreases the duration of the matter dominated era (and at fix $h$, a smaller $\Omega_m$/larger $\Omega_{\Lambda}$), implying a shift of the power spectrum towards large scales, i.e. towards small wavenumbers. 
Second, DCDM models involve a larger ratio of $\omega_{\rm b}/\omega_{\rm cdm}$ compared to the $\Lambda$CDM model due to the decay. 
Both effects manifests as a strong suppression of the small-scale power spectrum, and the latter effect leads to an additional modulation of the BAO amplitude visible as wiggles in Figs. \ref{fig:DR_linear} and \ref{fig:pybird_dcdm_dr}. 
Moreover, we note that the non-linear matter power spectrum shows a stronger scale-dependent suppression compared to the linear power spectrum at $k \gtrsim 0.1 \ h\,\text{Mpc}^{-1}$. There is an intuitive explanation as to why the non-linear power spectrum is further suppressed, very similarly to what happens for standard neutrinos or warm dark matter, as reviewed e.g. in \cite{Lesgourgues:2013sjj}. In general, non-linear growth is faster than the linear growth, and the impact of non-linearities is typically to enhance the power spectrum (this is famously the case in $\Lambda$CDM). In the DCDM case, modes that are suppressed will enter the non-linear regime later, and therefore start experiencing their enhanced growth due to non-linearities later. This delay leads to a further suppression of the power spectrum compared to $\Lambda$CDM when non-linear effects are included. We checked that the amplitude of the deviation from scale-independent suppression at $k \gtrsim 0.1 \ h\,\text{Mpc}^{-1}$ is tied to the value of the effective dark matter sound speed $c_{s}$, and can vary of a few $\%$ for $c_{s} \in [1,5] \ k_{\rm nl}^2\cdot({\rm Mpc}/h)^2$, where $k_{\rm nl}$ corresponds to the non-linear scale and determines the cut-off scale of the theory. 
On the other hand, the power suppression gets less strong with larger $k$ in the monopole of the galaxy power spectrum, an effect indicating an additional degeneracy with other EFT parameters. 
Finally, and as expected, deviations with respect to $\Lambda$CDM increases as  $\tau$ decreases and/or $f_{\rm dcdm}$ increases for the monopole as well as for the quadrupole.

\subsubsection{Preliminary study}
\label{sec:Preliminary_study_dr}

\begin{figure*}
    \centering
\includegraphics[trim=1cm 0cm 3cm 1cm, clip=true,width=2\columnwidth]{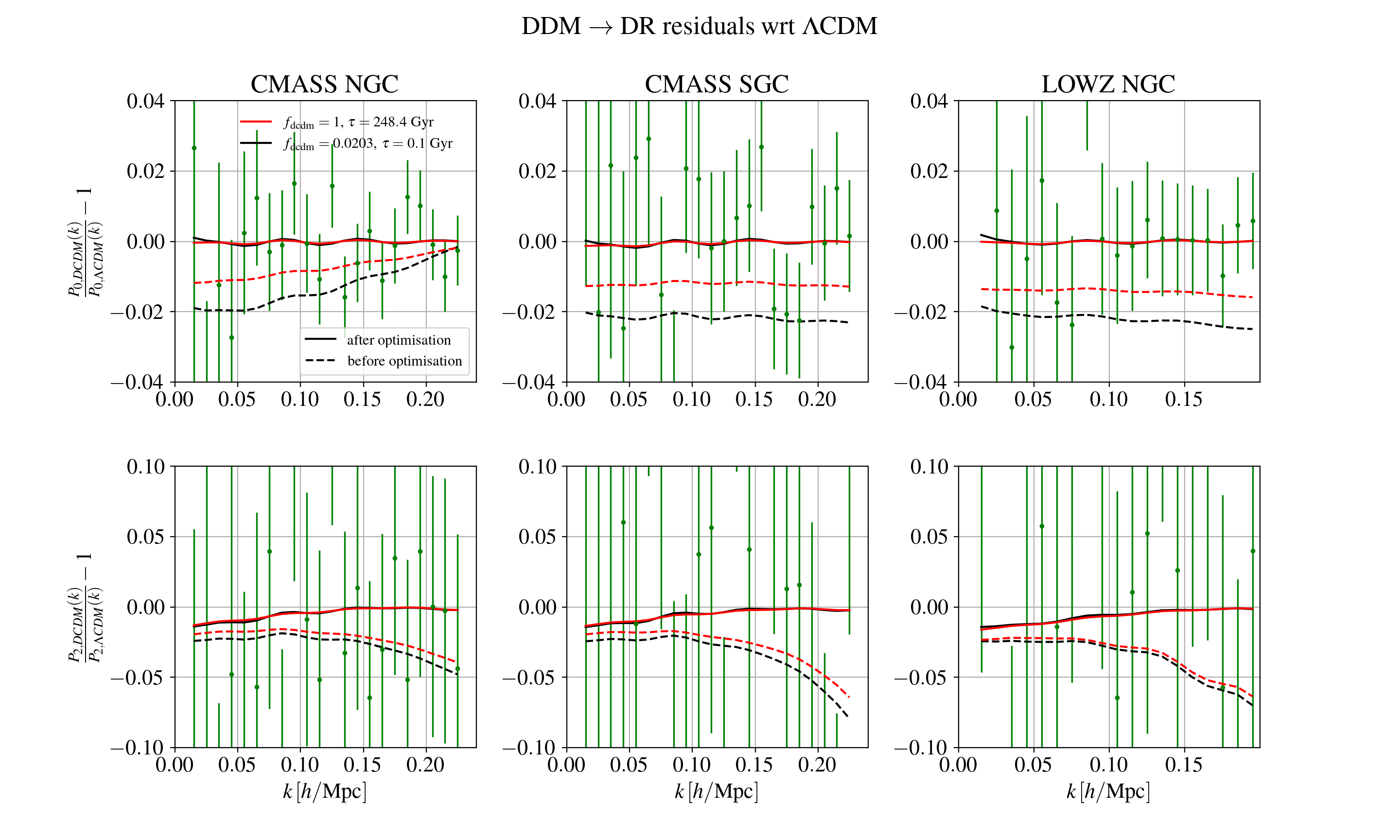}
\caption{Residuals of the monopole and the quadrupole of our DCDM $\to$ DR preliminary study with respect to $\Lambda$CDM model (EFTofBOSS analysis in Tab. \ref{tab:LCDM}) for the three sky-cuts of the EFTofBOSS data. For the solid lines we optimized the EFT nuisance parameters, while for the dotted lines we set the EFT nuisance parameters to those of the $\Lambda$CDM (EFTofBOSS analysis in Tab. \ref{tab:LCDM}).}
    \label{fig:forecast_dr_without_BOSS}
\end{figure*}

To gauge the impact of using the EFTofBOSS data in our analyses of the DCDM $\to$ DR model, we first perform a preliminary study in which we consider a set of DCDM parameters laying at the 95\% C.L.\footnote{From here on, we quote one-sided bounds at 2$\sigma$ (95 \% C.L.) and two-sided bounds at 1$\sigma$ (68\% C.L.).} derived from {\it Planck} data, and compute the $\chi^2$ of the EFTofBOSS data after optimising the EFT nuisance parameters. The goal is to check the extent to which EFT nuisance parameters can lead to effects degenerate with those of the DCDM with a quick analysis.
We set all $\Lambda$CDM parameters to their best-fit values from the analysis of  `{\it Planck} + Pantheon + EFTofBOSS + Ext-BAO' (see Sec.~\ref{sec:DataAnalysis}). We perform two analyses: (i) we set $\tau = 0.1$ Gyr and take the upper bound on $f_{\rm dcdm}$ from our `{\it Planck} + Pantheon + Ext-BAO (no Ly-$\alpha$)', i.e. $f_{\rm dcdm} = 0.0203$ (see Tab. \ref{tab:DR_2sigma_limit}), and (ii) we set $f_{\rm dcdm}=1$ (i.e. all the dark matter decays), while we take the lower bound of $\tau$ from our `{\it Planck} + Pantheon + Ext-BAO (no Ly-$\alpha$)' analysis, i.e. $\tau = 248.4$ Gyr (see Tab. \ref{tab:DR_2sigma_limit}). 
We show in Tab.~\ref{tab:DRnuisance_preliminary} the $\chi^2$ associated to the EFTofBOSS data, and we plot in Fig~\ref{fig:forecast_dr_without_BOSS}, using the PyBird code, the residuals (with respect to $\Lambda$CDM from the `{\it Planck} + Pantheon + EFTofBOSS + Ext-BAO' analysis) of these studies. 
To gauge the impact of EFT nuisance parameters, in this latter figure, we show residuals with and without the optimization procedure (in the latter case, we simply set the EFT nuisance parameters to those of $\Lambda$CDM). 
This preliminary study allows us to highlight two important points. 
Firstly, the optimization procedure has washed out the suppression due to decay, which implies that the effect of the EFT nuisance parameters are (at least partly) degenerate with that of the decay.
Secondly (and consequently), for these two analyses where we have chosen DCDM parameters that are excluded at 95\% C.L., we obtain a $\chi^2$ very close to that of the $\Lambda$CDM best-fit model of the full analysis, suggesting that EFTofBOSS data may not provide strong additional constraints to this model. 
Naturally, it does not prevent the model to potentially yield an improved fit over $\Lambda$CDM once all (cosmological and nuisance) parameters are optimized against the data, and we will check our naive results against a full analysis in Sec.~\ref{sec:DataAnalysis}.

\begin{table*}[t]
\begin{tabular}{|l|c|c|}
 \hline
 Parameter & $f_{\rm dcdm} = 0.0203$ \& $\tau=0.1$ Gyr & $f_{\rm dcdm}=1$ \& $\tau=248.4$ Gyr \\
\hline
$\chi^2_{\text{CMASS NGC}}$ & 41.3 & 40.7 \\
$\chi^2_{\text{CMASS SGC}}$ & 43.9 & 44.0 \\
$\chi^2_{\text{LOWZ NGC}}$ & 33.4 & 33.6 \\
\hline
$\chi^2_{\text{EFTofBOSS}}$ & 118.6 & 118.3 \\
$\footnotesize{\chi^2_{\rm min} (\text{DCDM}) -\chi^2_{\rm min} (\Lambda\text{CDM})} $ & +0.8  & +0.5\\
\hline
\end{tabular}
\caption{$\chi^2$ of each sky-cut of the EFTofBOSS data set for our DCDM $\to$ DR preliminary study. We also indicated the $\Delta\chi^2$ with respect to the analogous $\Lambda$CDM best-fit model (EFTofBOSS analysis in Tab. \ref{tab:chi2_EFT}).}
\label{tab:DRnuisance_preliminary}
\end{table*}

\subsection{Warm Dark Matter decay products (DCDM $\to$ WDM+DR model)}

\subsubsection{Presentation of the model}
\label{sec:WDM_review}
We now turn to a DCDM model where the entirety of the DM sector is considered unstable (i.e. $f_{\rm dcdm}=1$ in the language of the first model), decaying into dark radiation and a massive particle, which will act as WDM. 
As before, we assume the decay products do not interact with the standard model particles.
The DCDM sector is now described by the DCDM lifetime  $\tau$, and the fraction $\varepsilon$ of rest-mass energy carried away by the massless particle  given by \cite{Blackadder_2014}
\begin{equation}
    \varepsilon = \frac{1}{2}\left(1-\frac{m^2_{\text{wdm}}}{m^2_{\text{dcdm}}}\right),
\end{equation}
where $m_{\text{dcdm}}$ and $m_{\text{wdm}}$ are the mother and daughter particle masses respectively. The accurate computation of the cosmological impact of the DCDM sector requires to follow the evolution of the phase space distribution of the warm particle produced during the decay. The full set of equations is described in Ref.~\cite{Aoyama_2014,abellan_2021}. 
We summarize here the sets of equations describing the evolution of the background energy densities  of the dark components, as well as the linear perturbations in a fluid approximation, valid well within the horizon.

\

The background energy densities evolve as follows \cite{Aoyama_2014}:
\begin{align}
   \dot{\bar{\rho}}_{\text{dcdm}} + 3 \mathcal{H}\bar{\rho}_{\text{dcdm}} &= -a\Gamma \bar{\rho}_{\text{dcdm}}, \\
     \dot{\bar{\rho}}_{\text{wdm}} + 3(1+w) \mathcal{H}\bar{\rho}_{\text{wdm}} &= (1-\varepsilon)a\Gamma \bar{\rho}_{\text{dcdm}}, \\
\dot{\bar{\rho}}_{\text{dr}} + 4 \mathcal{H}\bar{\rho}_{\text{dr}} &= \varepsilon\Gamma
a \bar{\rho}_{\text{dcdm}}\,
\end{align}
where $w=\bar{P}_{\text{wdm}}/\bar{\rho}_{\text{wdm}}$ is the equation of state of the massive daughter particle. 
 In the limit of large $\tau$ or small $\varepsilon$, one recovers the $\Lambda$CDM model, while setting $\varepsilon=1/2$ leads to a decay solely into massless particles.

\

In the synchronous gauge comoving with the DCDM fluid, the linear perturbation equations for the parent particle and DR daughter is still given by Eq. \eqref{delta_DCDM_evolution} and Eqs.  \ref{delta_DR_evolution_1}-\ref{l3_DR_evolution_1}, respectively. However, the quantity $r_{\rm dr}$ now satisfies 
\begin{equation}
\dot{r}_\dr = a \varepsilon \Gamma 
(\bar{\rho}_\dcdm/\bar{\rho}_\dr) r_\dr, 
\end{equation}
where the parameter $\varepsilon$ now affects the amount of energy transferred to the DR.
Regarding the WDM linear perturbations, it is unfortunately not possible to integrate out the dependency on momenta as it is done for the DR species. In general one has to follow the evolution of the full phase-space distribution, which becomes very computationally demanding (see Ref.~\cite{abellan_2021} for the expression of the full Boltzmann hierarchy). Nevertheless, it was shown in Ref.~\cite{abellan_2021} that, well within the horizon, the dynamics of the WDM perturbations can be well approximated by the following set of fluid equations:
\begin{align}
    \dot{\delta}_{\text{wdm}} &= -3\mathcal{H}(c_s^2-\omega)\delta_{\text{wdm}} - (1+\omega)\left(\theta_{\text{wdm}} + \frac{\dot{h}}{2}\right) \nonumber \\ 
    &+ (1-\varepsilon)a\Gamma \frac{\bar{\rho}_{\text{dcdm}}}{\bar{\rho}_{\text{wdm}}}(\delta_{\text{dcdm}}-\delta_{\text{wdm}}) \label{delta_WDM_evolution},\\
    \dot{\theta}_{\text{wdm}} &= -\mathcal{H}(1-3c_g^2)\theta_{\text{wdm}} + \frac{c_s^2}{1+\omega}k^2\delta_{\text{wdm}} - k^2\sigma_{\text{wdm}} \nonumber\\
    &- (1-\varepsilon)a\Gamma\frac{1+c_g^2}{1+\omega} \frac{\bar{\rho}_{\text{dcdm}}}{\bar{\rho}_{\text{wdm}}}\theta_{\text{wdm}}, \label{theta_WDM_evolution}
\end{align}
where $c_s$ is the WDM sound speed in the synchronous gauge, i.e. $c_s^2 = \delta P_{\text{wdm}}/ \delta \rho_{\text{wdm}}$, and $c_g$ is the WDM adiabatic sound speed, i.e. $c_g^2 = \dot{\bar{P}}_{\text{wdm}}/\dot{\bar{\rho}}_{\text{wdm}}$, which one can write in the following form:
\begin{align}
    c_g^2 &= w\left(  5 - \frac{\mathfrak{p}_{\text{wdm}}}{\bar{P}_{\text{wdm}}} - \frac{\bar{\rho}_{\text{dcdm}}}{\bar{\rho}_{\text{wdm}}}  \frac{a\Gamma}{3w\mathcal{H}} \frac{\varepsilon^2}{1-\varepsilon} \right) \nonumber \\
    &\times \left[  3(1+w) - \frac{\bar{\rho}_{\text{dcdm}}}{\bar{\rho}_{\text{wdm}}} \frac{a\Gamma}{\mathcal{H}}(1-\varepsilon) \right]^{-1}.
\end{align}
In this latter equation, $\mathfrak{p}_{\text{wdm}}$ is the pseudo-pressure (introduced in the context of the fluid equations for massive neutrinos \cite{Lesgourgues:2011rh}), which corresponds to a higher momenta integral of the WDM homogeneous phase  space distribution, reducing to the standard pressure in the relativistic limit.
Solving the fluid equations requires specifying the sound speed $c_s$, which was found to be well described by the following formula:
\begin{equation}
    c_s^2(k,\tau)=c_g^2[1+0.2\times(1-2\varepsilon)\sqrt{k/k_{\rm fs}}]\,
\end{equation}
where the free-streaming scale $k_{\rm fs}$ of the WDM is computed as:
\begin{equation}
\label{eq:kfs}
    k_{\text{fs}}(\tau) = \sqrt{\frac{3}{2}}\frac{\mathcal{H}(\tau)}{c_g(\tau)}\,.
\end{equation}
 The free-streaming scale corresponds to the scale at which pressure (coming from the `velocity kick' received during the decay process) suppresses  perturbations of the WDM compared to those of the DCDM. In other word, on scales $k<k_{\rm fs}$, one has $\delta_{\text{wdm}} = \delta_{\text{dcdm}}$, while on scale $k>k_{\rm fs}$ the WDM perturbations are suppressed and exhibit oscillations over time.

\

To obtain the linear CMB and matter power spectrum, we make use of  an extension of the CLASS code\footnote{\url{https://github.com/PoulinV/class_decays}} described in Ref.~\cite{abellan_2021}, and we determine the non-linear galaxy power spectrum using the PyBird code. 
We have argued in previous section and in App.~\ref{sec:app_EFT_vs_Nbody}, through direct comparison with N-body simulations, that PyBird can safely be used to describe DM decays with massless decay products. Unfortunately, we do not have access to such N-body simulations in the case of massive decay products.
 A priori, the problem is not the decay per se (as we have seen for the massless decay products). Rather, contrarily to the case of massless daugthers, the massive daughter may develop perturbations whose contribution to the total matter power spectrum can be highly non-trivial. In App.~\ref{sec:app_EFT_WDM}, following Refs~\cite{senatore2017neutrinos,senatore2019neutrinos}, which treated the similar case of massive neutrinos, we argue that the corrections to the EFTofLSS necessary to fully capture the model-specific effects can be neglected for most of the parameter space of interest, as the fractional contribution of the WDM to the DM density is small (in particular for the best-fit model that we derive), or the free-streaming scale exceeds the scale-cut considered in the analysis.

\subsubsection{The non-linear power spectrum}

\begin{figure*}
    \centering
    \includegraphics[trim=2cm 0cm 3cm 11cm, clip=true, width=2.01\columnwidth]{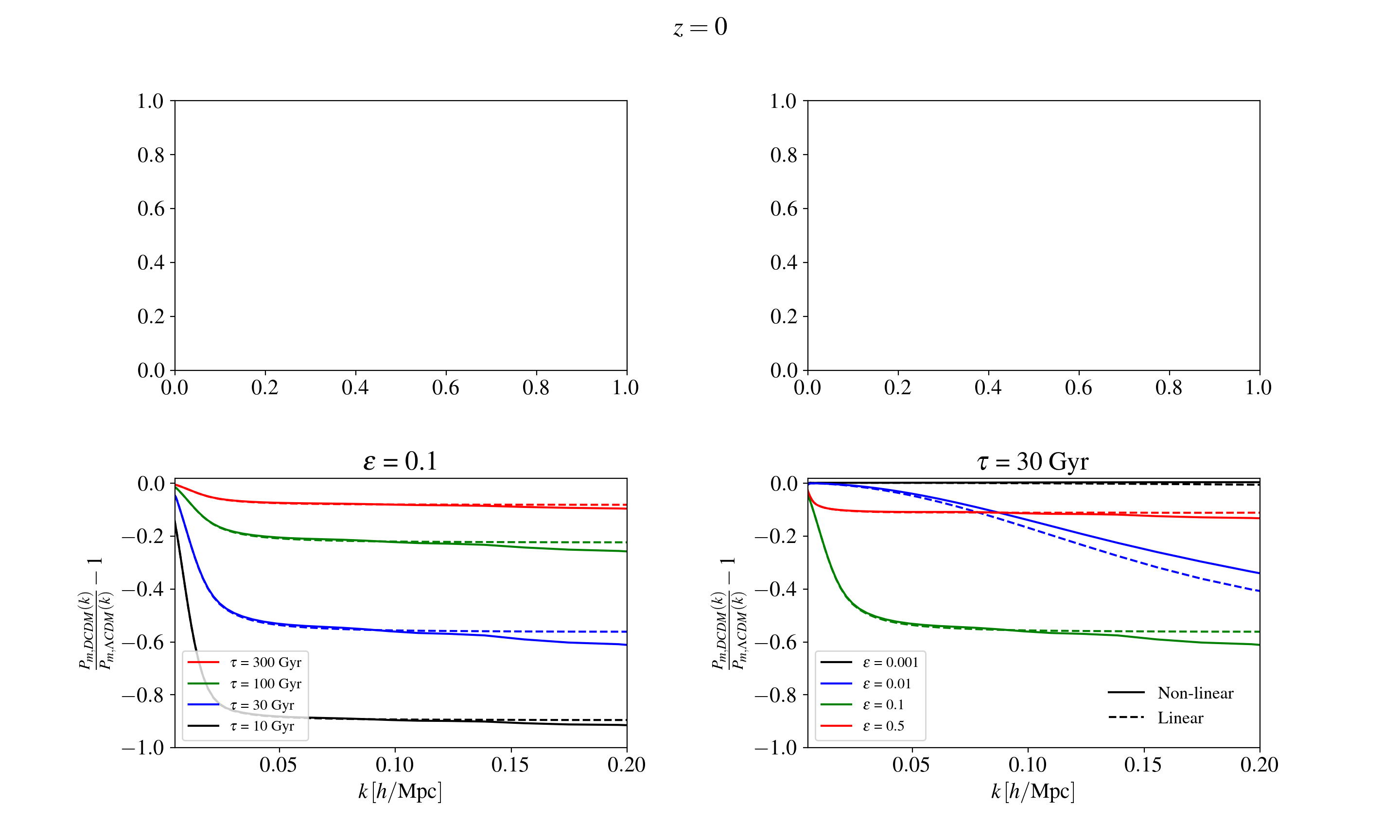}
    \caption{\textit{Left} - Residuals of the linear (dashed lines) and non-linear matter power spectrum (solid lines) for $\varepsilon$ set to 0.1 and $\tau =$ 10, 30, 100, 300 Gyr. Residuals are taken with respect to the $\Lambda$CDM model at $z=0$. \textit{Right} - The same, but this time $\tau$ is set to 30 Gyr and $\varepsilon = $ 0.001, 0.01, 0.1, 0.5.}
    \label{fig:WDM_linear}
    \centering
    \includegraphics[trim=2cm 0cm 3cm 0cm, clip=true,width=2.01\columnwidth]{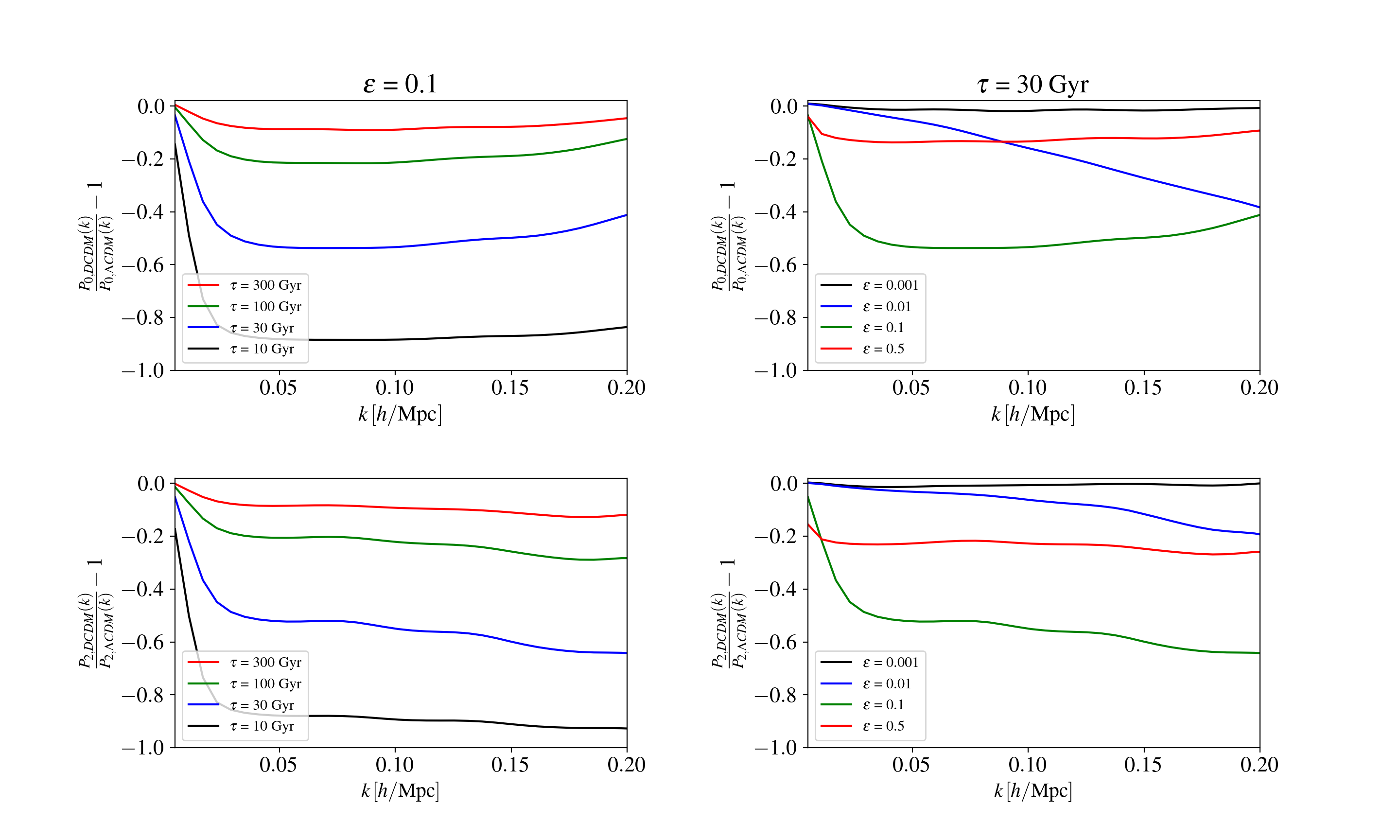}
    \caption{\textit{Left} - Residuals of the monopole and the quadrupole of the galaxy power spectrum for $\varepsilon$ set to 0.1 and $\tau =$ 10, 30, 100, 300 Gyr. Residuals are taken with respect to the $\Lambda$CDM model at $z=0$. \textit{Right} - The same, but this time $\tau$ is set to 30 Gyr and $\varepsilon = $ 0.001, 0.01, 0.1, 0.5.}
    \label{fig:PyBird_dcdm_wdm}
\end{figure*}

We plot in Fig. \ref{fig:WDM_linear} the residuals of the non-linear matter power spectra of the DCDM $\to$ WDM+DR model with respect to that of the $\Lambda$CDM model at $z=0$. We also represent the associated linear matter power spectra obtained from the CLASS code, exactly as in Fig. 4 of Ref. \cite{abellan_2021}. In addition, we plot in Fig. \ref{fig:PyBird_dcdm_wdm} the residuals of the monopole and quadrupole of the galaxy power spectra of this model. In these figures, the cosmological parameters are taken from the DCDM $\to$ WDM+DR best-fit model of Ref. \cite{Abellan:2020pmw}, while the nuisance parameters are set as in Figs. \ref{fig:DR_linear} and \ref{fig:pybird_dcdm_dr}. In the left panels, we fix $\varepsilon=0.1$ and vary $\tau\in [10,300]$ Gyr, while in the right panel  we fix $\tau = 30$ Gyr and vary $\varepsilon\in[0.001,0.5]$. 

\

As for the case of the DCDM $\to$ DR model, we obtain a very similar behaviour between the linear matter power spectrum and the monopole of the galaxy power spectrum, except for a mild monotonic reduction of the power suppression at larger $k$'s in the monopole of the galaxy power spectrum (due to the choice of EFT parameters, this reduction of the suppression may change for different values). The presence of a warm dark matter component which does not cluster on small scales suppresses the matter power spectrum as well as the galaxy power spectrum, and $\tau$ -- which sets the abundance of the WDM species today -- controls  the amplitude of the power suppression, while $\varepsilon$ controls the cutoff scale.  One can see in Fig. \ref{fig:PyBird_dcdm_wdm} that the suppression of the galaxy spectrum increases as $\tau$ decreases (left panel), while the suppression starts to occur on larger scales as $\varepsilon$ increases (right panel). Once $\varepsilon =0.5$, the free-streaming scale $k_{\text{fs}}$ becomes equivalent to the Hubble horizon, and the effects become identical to that of the DCDM $\to$ DR model presented before. Note that because of the effect of the WDM, the $\varepsilon=0.1$ case has a stronger suppression than the $\varepsilon=0.5$ (pure dark radiation) case.
Moreover, we find (see Fig. \ref{fig:WDM_linear}) that the non-linear correction slightly modulates the slope of the power suppression compared to the linear matter power spectrum. It always leads to a stronger suppression than the linear one at large enough $k$ (for $\varepsilon \gtrsim 0.1$, the modulation occurs at $k \gtrsim 0.1 \ h\,\text{Mpc}^{-1}$). However, for smaller $\varepsilon$ (see the $\varepsilon=0.01$ case for example), the modulation can appear as a milder power suppression compared to the linear one in the range of validity of the EFT at one-loop order.

\subsubsection{Preliminary study}
\label{sec:Preliminary_study_wdm}

\begin{table}[t]
\begin{tabular}{|l|c|}
 \hline
 Parameter & Best-fit  \\
\hline
$\chi^2_{\text{CMASS NGC}}$ & 41.2 \\
$\chi^2_{\text{CMASS SGC}}$ & 44.5\\
$\chi^2_{\text{LOWZ NGC}}$ & 34.4 \\
\hline
$\chi^2_{\text{EFTofBOSS}}$ & 120.1\\
$\footnotesize{\chi^2_{\rm min} (\text{DCDM}) -\chi^2_{\rm min} (\Lambda\text{CDM})} $ &  +3.1\\
\hline
\end{tabular}
\caption{$\chi^2$ of each sky-cut of the EFTofBOSS data set for our DCDM $\to$ WDM+DR preliminary study. We also indicated the $\Delta\chi^2$ with respect to the analogous $\Lambda$CDM best-fit model (EFTofBOSS + $S_8$ analysis in Tab. \ref{tab:chi2_EFT}).}
\label{tab:WDMnuisance_preliminary}
\end{table}

\begin{figure*}
    \centering
\includegraphics[trim=1cm 0cm 3cm 1cm, clip=true, width=2\columnwidth]{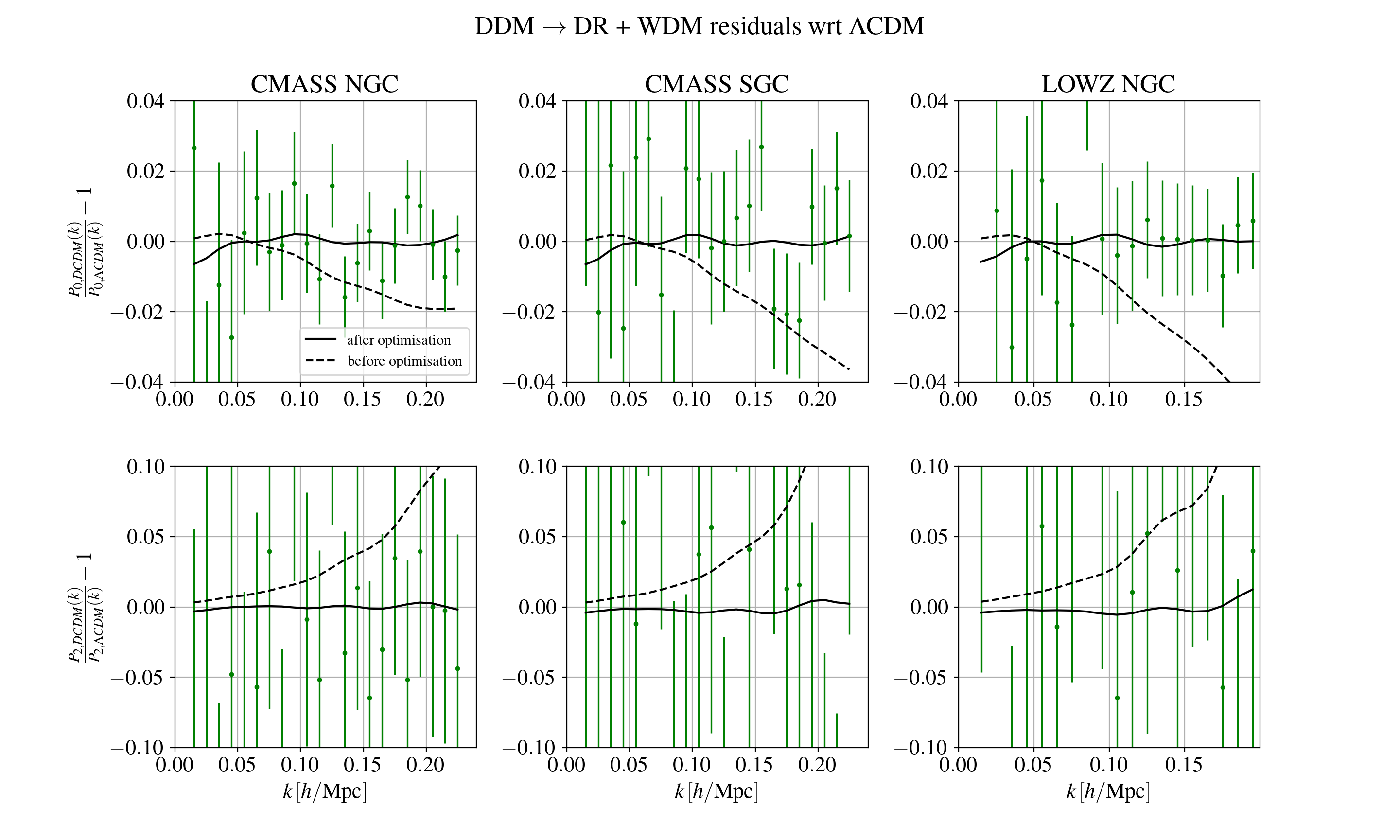}
\caption{Residuals of the monopole and the quadrupole of our DCDM $\to$ WDM+DR preliminary study with respect to $\Lambda$CDM model (EFTofBOSS analysis in Tab. \ref{tab:LCDM}) for the three sky-cuts of the EFTofBOSS data. For the solid lines we optimized the EFT nuisance parameters, while for the dotted lines we set the EFT nuisance parameters to those of the $\Lambda$CDM (EFTofBOSS analysis in Tab. \ref{tab:LCDM}).}
    \label{fig:forecast_wdm_without_BOSS}
\end{figure*}

Similarly to the case of the DCDM $\to$ DR model, we perform a preliminary study to test whether the EFTofBOSS data can further constrain the DCDM $\to$ WDM+DR model that resolves the $S_8$ tension. We fix cosmological parameters\footnote{The analysis performed in Refs.~\cite{Abellan:2020pmw,abellan_2021} made use of a $S_8$ prior that includes information from BOSS \cite{Heymans:2020gsg}. For consistency and to avoid double counting information, we re-performed the analysis (see Sec. \ref{sec:DataAnalysis}) with a prior derived from KiDS-1000 data alone.} to those obtained from the joint analysis of {\it Planck} data, Pantheon SN1a data, a compilation of BAO data and the $S_8$ measurements by KiDS-1000 \cite{KiDS:2020suj}.
 We optimize the EFT nuisance parameters of the galaxy power spectrum to check the extent to which they can lead to effects degenerate with those of the DCDM. We show in Tab.~\ref{tab:WDMnuisance_preliminary} the $\chi^2$ associated to the EFTofBOSS data, while in Fig. \ref{fig:forecast_wdm_without_BOSS}, using the PyBird code, we plot the residuals with respect to the best-fit $\Lambda$CDM model from the analysis of `{\it Planck} + Pantheon + EFTofBOSS + Ext-BAO' (see Sec.~\ref{sec:DataAnalysis}). In this figure, we represent residuals with and without the EFT optimization procedure (in the latter case, we simply set the EFT nuisance parameters to those of $\Lambda$CDM). As before, one can see that the effects of the DCDM are strongly reduced once EFT nuisance parameters are optimized, suggesting a strong degeneracy between the DCDM and the EFT parameters. Nevertheless, for this preliminary study, the $\chi^2$ is degraded by $+3.1$ compared to the best-fit $\chi^2$ obtained in the $\Lambda$CDM model for the full analysis. Contrary to the preliminary study of the DCDM $\to$ DR model for which we obtained a $\chi^2$ close to that of the $\Lambda$CDM model, we anticipate that the EFTofBOSS data can provide additional constraining power to this model.

\section{\label{sec:DataAnalysis}A comprehensive MCMC analysis of the DCDM models}

\subsection{Data and method}

We now perform a Monte Carlo Markov Chain (MCMC) analyses, confronting these two DCDM models with recent cosmological observations. 
To do so, we make use of the MontePython-v3 code \cite{Brinckmann:2018cvx,Audren:2012wb} interfaced with our modified CLASS version. 
We perform various analyses from a combination of the following data sets:
\begin{itemize}
    \item {\bf Planck:} The low-$l$ CMB TT, EE, and the high-$l$ TT, TE, EE data, as well as the gravitational lensing potential reconstruction from {\it Planck} 2018 \cite{Aghanim:2018eyx,Planck:2018lbu}.
    \item {\bf Pantheon:} The Pantheon SNIa catalogue, spanning redshifts $0.01 < z < 2.3$ \cite{Scolnic:2017caz}.
    \item {\bf Ext-BAO:} The BAO measurements from 6dFGS at $z = 0.106$, SDSS DR7 at $z = 0.15$ \cite{Beutler:2011hx,Ross:2014qpa,Alam:2016hwk}, and the joint constraints from eBOSS DR14 Ly-$\alpha$ absorption auto-correlation at $z = 2.34$ and cross-correlation with quasars at $z = 2.35$ \cite{Agathe:2019vsu, Blomqvist:2019rah}.
    \item {\bf BOSS BAO/$\bm{f\sigma_8}$:} The measurements of the BAO and the redshift space distortion $f\sigma_8(z)$ from the CMASS and LOWZ galaxy samples of BOSS DR12 at $z = 0.38$, 0.51, and 0.61 \cite{Alam:2016hwk}.
    \item {\bf$\bm{S_8}$:} The KIDS-1000 cosmic shear measurement of $S_8=0.759^{+0.024}_{-0.021}$, modeled as a a split-normal likelihood \cite{KiDS:2020suj}.
    \item {\bf EFTofBOSS:} The CMASS and LOWZ data sets of the EFTofBOSS data (see Sec. \ref{sec:EFT}).
\end{itemize}
Our analyses always include {\it Planck}, Pantheon and Ext-BAO data. However, we quantify the impact of EFTofBOSS data and the $S_8$ prior by performing analyses with and without these data. When {\it not} including the EFTofBOSS data, we make use of the conventional BOSS BAO/$f\sigma_8$ data.
We use \textit{Planck} conventions for the treatment of neutrinos and include two massless and one massive species with $m_{\nu} = 0.06$ eV~\cite{Aghanim:2018eyx}. 
We impose a large flat prior on the dimensionless baryon energy density $\omega_b$, the Hubble parameter today $H_0$, the logarithm of the variance of curvature perturbations centered around the pivot scale $k_p = 0.05$ Mpc$^{-1}$ (according to the {\it Planck} convention), $\ln(10^{10}\mathcal{A}_s)$, the scalar spectral index $n_s$, and the re-ionization optical depth $\tau_{\rm reio}$. 
We assume our MCMC chains to be converged when the Gelman-Rubin criterion $R-1 < 0.05$ \cite{Gelman_1992}. Finally, we extract the best-fit parameters from the procedure highlighted in appendix of Ref.~\cite{Schoneberg:2021qvd}.

 \begin{table*}
\begin{tabular}{|l|c|c|}
 \hline
\multicolumn{3}{|c|}{$\Lambda$CDM} \\
 \hline
 Parameter & w/ EFTofBOSS  & w/ EFTofBOSS + $S_8$ \\
 \hline
$100 \ \omega_{\rm b} $                                    & ~~~~~~~~~~   $2.242(2.245)_{-0.015}^{+0.014}$     ~~~~~~~~~~    &  ~~~~~~~~~~  $2.247(2.248)\pm0.014$ ~~~~~~~~~~   \\
$\omega_{\text{cdm}}$ & $0.1191(0.1191)\pm0.00095$ & $0.1184(0.1184)\pm+0.00089$\\
$H_0 /[{\rm km/s/Mpc}]$ & $67.76(67.80)_{-0.44}^{+0.42}$ & $68.05(68.07)\pm0.41$ \\
$\text{ln}(10^{10} A_s)$                             & $3.048(3.049)_{-0.016}^{+0.015}$  & $3.043(3.043)_{-0.016}^{+0.015}$ \\
$n_s$                                                & $0.9666(0.9676)\pm0.0039$     & $0.9680(0.9687)\pm0.0039$ \\
$\tau_{\rm reio}$								     & $0.0571(0.0574)_{-0.0085}^{+0.0075}$      & $0.0555(0.0549)_{-0.0078}^{+0.0077}$ \\
\hline
$\Omega_{\rm m}$								     & $0.3098(0.3093)_{-0.0058}^{+0.0057}$     & $0.3057(0.3055)\pm0.0053$ \\
$\sigma_8$										         & $0.8097(0.8102)_{-0.0065}^{+0.0063}$      & $0.8056(0.8055)\pm0.0062$ \\
$S_8$ & $0.82(0.82)\pm0.01$ & $0.813(0.813)_{-0.0096}^{+0.0094}$ \\
\hline
$\chi^2_{\rm min}$ & 3927.0   & 	3933.0		\\
\hline
$Q_{\rm DMAP}\equiv\sqrt{\chi^2_{\rm min}({\rm w/}~S_8)-\chi^2_{\rm min}({\rm w/o}~S_8)} $ & \multicolumn{2}{|c|}{2.4$\sigma$}\\
\hline
\end{tabular}
\caption{The mean (best-fit) $\pm 1\sigma$ errors of the cosmological parameters from our `{\it Planck} + Pantheon + EFTofBOSS + Ext-BAO' and `{\it Planck} + Pantheon + EFTofBOSS + Ext-BAO + $S_8$' analyses for the $\Lambda$CDM model. For each data set we also report its best-fit $\chi^2$.}
\label{tab:LCDM}
\end{table*}

\begin{table*}
\begin{tabular}{|l|c|c|}
 \hline

\multicolumn{3}{|c|}{DCDM$\to$DR} \\

 \hline
 Parameter & w/ EFTofBOSS  & w/ EFTofBOSS + $S_8$ \\
 \hline
 $\Gamma/[{\rm  Gyr}^{-1}]$  & ~~~~~~~~~~  	unconstrained (4.8) ~~~~~~~~~~  	 &  ~~~~~~~~~~      unconstrained (5.8)     ~~~~~~~~~~  \\
$f_{\rm dcdm}$					     &  $<0.0216(1.62\cdot 10^{-4})$   & $<0.0242(1.67 \cdot 10^{-4})$ \\
\hline
$100 \ \omega_{\rm b} $                               &  $2.236(2.244) \pm 0.015$     & $2.241(2.248)_{-0.015}^{+0.016}$ \\
$\omega_{\text{cdm}}$ & $0.1187(0.1191)\pm 0.0010$ & $0.1180(0.1184)_{-0.00093}^{+0.001}$\\
$H_0 /[{\rm km/s/Mpc}]$ & $67.98(67.77)_{-0.48}^{+0.46}$ & $68.30(68.10)_{-0.47}^{+0.44}$\\
$\text{ln}(10^{10} A_s)$                             &  $3.051(3.049)_{-0.016}^{+0.015}$ & $3.047(3.045)_{-0.016}^{+0.015}$ \\
$n_s$                                                &   $0.9650(0.9671)_{-0.004}^{+0.0042}$    &  $0.9660(0.9687)_{-0.0043}^{+0.0044}$\\
$\tau_{\rm reio}$								     &   $0.0577(0.0572)_{-0.0079}^{+0.0073}$     &  $0.0562(0.0557)_{-0.0077}^{+0.0074}$ \\
\hline
$\Omega_{\rm m}$								     &  $0.3069(0.3097) \pm 0.0061$    & $0.3026(0.3050)_{-0.0057}^{+0.0059}$ \\
$\sigma_8$							       &    $0.8110(0.8101)_{-0.0066}^{+0.0063}$   & $0.8071(0.8061)_{-0.0063}^{+0.0062}$ \\
$S_8$ & $0.82(0.82)\pm 0.01$ & $0.811(0.813)_{-0.0095}^{+0.0097}$ \\

\hline
$\chi^2_{\rm min}$ &  3927.0  & 	3933.0		\\
\hline
$\footnotesize{\chi^2_{\rm min} (\text{DCDM}) -\chi^2_{\rm min} (\Lambda\text{CDM})} $ & 0.0 & 0.0 \\
\hline
$Q_{\rm DMAP}\equiv\sqrt{\chi^2_{\rm min}({\rm w/}~S_8)-\chi^2_{\rm min}({\rm w/o}~S_8)} $ & \multicolumn{2}{|c|}{2.4$\sigma$}\\
\hline
\end{tabular}
\caption{The mean (best-fit) $\pm 1\sigma$ errors of the cosmological parameters from our `{\it Planck} + Pantheon + EFTofBOSS + Ext-BAO' and `{\it Planck} + Pantheon + EFTofBOSS + Ext-BAO + $S_8$' analyses for the DCDM $\to$ DR model. For each data set we also report its best-fit $\chi^2$, and the $\Delta\chi^2$ with respect to the analogous $\Lambda$CDM best-fit model.}
\label{tab:DR}
\end{table*}

\begin{table*}
\addtolength{\tabcolsep}{+3pt}
\begin{tabular}{|l|c|c|}
 \hline
\multicolumn{3}{|c|}{DCDM $\to$ DR} \\
\hline
Data sets & $f_{\rm dcdm}$ & $\tau$ (for $f_{\rm dcdm}=1$)\\
\hline
{\it Planck} & $< 0.0205$ & $>$246.3 Gyr\\
\hline
{\it Planck} + Pantheon + Ext-BAO (no Ly-$\alpha$) & $< 0.0203$ & $>$248.4 Gyr\\
\hline
{\it Planck} + Pantheon + BOSS BAO/$f\sigma_8$ + Ext-BAO (no Ly-$\alpha$) & $<0.0190$ & $>$260.4 Gyr\\
\hline
{\it Planck} + Pantheon + BOSS BAO/$f\sigma_8$ + Ext-BAO& $<$0.0219 & $>$250.0 Gyr \\
\hline
{\it Planck} + Pantheon + EFTofBOSS + Ext-BAO & $<$0.0216 & $>$249.6 Gyr\\
\hline
\end{tabular}
\caption{The 95\% C.L. limit on $f_{\rm dcdm}$ for the standard DCDM $\to$ DR analysis, and the 95\% C.L. limit on $\tau$ for the DCDM $\to$ DR analysis where $f_{\rm dcdm}$ is fixed to the unit. Let us recall that `Ext-BAO' refers to the BAO measurements from 6dFGS, SDSS DR7, and the joint constraints from eBOSS DR14 Ly-$\alpha$ auto-correlation and cross-correlation. For some data sets we removed the Ly-$\alpha$ constraints ('no Ly-$\alpha$') to explicitly show its impact.}
\label{tab:DR_2sigma_limit}
\end{table*}
\subsection{Dark Radiation decay products}

\begin{figure*}
    \centering
    \includegraphics[width=1.3\columnwidth]{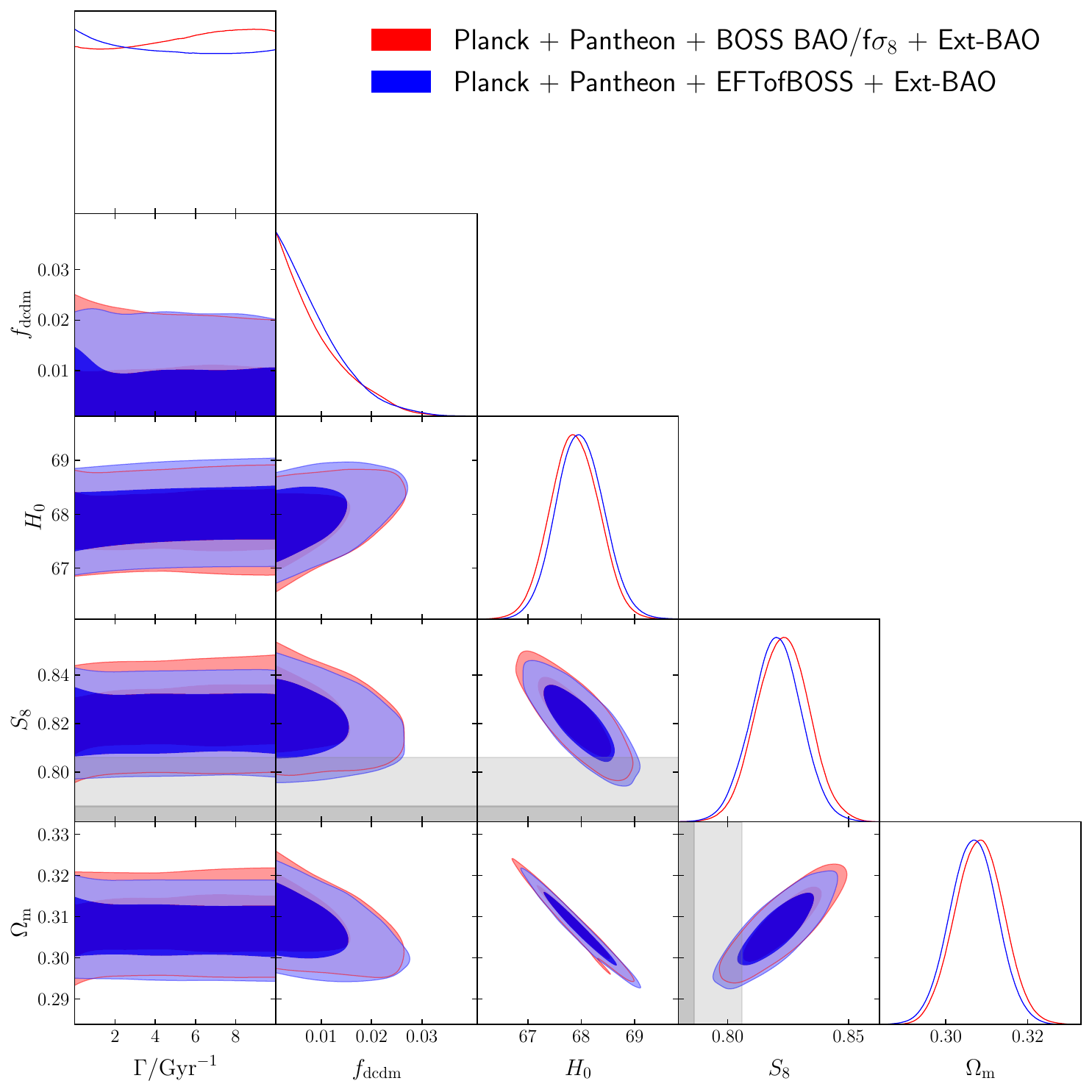}
        \caption{2D posterior distributions of the DCDM $\to$ DR model with and without the EFTofBOSS data set. The gray shaded bands refer to the joint $S_8$ measurement from KiDS-1000 + BOSS + 2dFLens \cite{Heymans:2020gsg}. }
    \label{fig:MCMCDR1}
\end{figure*}

Let us recall that in the case of the DCDM $\to$ DR model we have two additional parameters: $\Gamma =\tau^{-1}$, the decay rate of DCDM, and $f_{\rm dcdm}$, the fraction of DCDM with respect to the total DM. In the MCMC analyses, we impose flat priors on $\Gamma$ and $f$:
\begin{align*}
     0\le &~\Gamma /{\rm Gyr}^{-1} \le 10,\\
     0\le &~f_{\rm dcdm} \le 1.
\end{align*}
Our results for the analyses with and without $S_8$ prior are presented in Tab. \ref{tab:DR}, while the results of the analyses of $\Lambda$CDM against the same data sets are given in Tab.~\ref{tab:LCDM}. The $\chi^2$ of the EFTofBOSS data are reported in Tab. \ref{tab:chi2_EFT}. In Fig. \ref{fig:MCMCDR1}, we display the 1D and 2D posteriors of $\left\{\Gamma/{\rm  Gyr}^{-1},f_{\rm dcdm},H_0,S_8,\Omega_m\right\}$ for the DCDM $\to$ DR model with and without the EFTofBOSS data set. In App. \ref{sec:app_S8}, we represent the same figure, but this time with and without the $S_8$ prior (and with the EFTofBOSS data set for both). Without the $S_8$ prior, the $\Delta\chi^2$ with respect to $\Lambda$CDM is compatible with zero\footnote{The improvement is below the precision of ${\cal O}(0.1)$ that we estimated on the minimization, and we therefore simply quote $\Delta\chi^2=0.0$. 
Hereinafter, we follow the same approach when reporting other $\Delta\chi^2$.} (see Tab. \ref{tab:DR}), implying that the data does not favor the DCDM $\to$ DR model.
From Fig. \ref{fig:MCMCDR1},  one can see that the inclusion of the EFTofBOSS data does not improve the constraint on this model, which is consistent with the `naive' analysis presented in sec.~\ref{sec:Preliminary_study_dr}. 
Moreover, we show that when adding the $S_8$ prior, the $\Delta\chi^2$ with respect to $\Lambda$CDM is still compatible with zero (and the model does not provide a good fit to the $S_8$ prior) while the constraints on $\Gamma$ and $f_{\rm dcdm}$ are largely unaffected. We conclude (as in past studies) that this model does not resolve the $S_8$ tension.

\

To summarize our results, and present the most up-to-date constraints on DCDM with massless decay products, in Tab. \ref{tab:DR_2sigma_limit} we compare the 95\% C.L. limits obtained for $f_{\rm dcdm}$ and $\tau$ when successively adding data sets. 
To obtain the bounds on $f_{\rm dcdm}$ (in the `short-lived' regime), we marginalize over the parameter $\Gamma$ in the range described above.
On the other hand, to obtain the $\tau$ limits (in the `very long-lived' regime), we fix $f_{\rm dcdm}=1$ in our MCMC analyses, i.e. we assume that all DM decays. Note that, for $f_{\rm dcdm}
\to 1$, one can interpret our constraints as a limit on the ratio $\tau/f_{\rm dcdm}$, as discussed in Ref.~\cite{Poulin_2016}. From Tab. \ref{tab:DR_2sigma_limit}, one can deduce:
\begin{enumerate}
    \item[\textbullet]  The strongest constraints are obtained when considering `{\it Planck} + Pantheon + BOSS BAO/$f\sigma_8$ + Ext-BAO (no Ly-$\alpha$)'. In that case, we find $f_{\rm dcdm}<0.0190$ (in the short-lived regime), and $\tau/f_{\rm dcdm}>260.4$ Gyr (for $f_{\rm dcdm}\to 1$). 
    
    \item[\textbullet]   On the other hand the inclusion of  Ly-$\alpha$ BAO data slightly reduce the constraints. This is consistent with the fact that these data are compatible with $\Lambda$CDM only at the $1.7\sigma$ level \cite{Agathe:2019vsu, Blomqvist:2019rah}, favoring lower energy density at high-$z$ \cite{Poulin:2018zxs}. Additionally, we find that constraints with the EFTofBOSS data are the same as those with the standard redshift space distortion  $f\sigma_8$ information. Our fiducial constraints, including all data, is therefore $f_{\rm dcdm}<0.0216$, and $\tau/f>249.6$ Gyr. 

    \item[\textbullet] Our constraints are somewhat different than those derived in Ref. \cite{Nygaard:2020sow}, which considering {\it Planck} 2018 + BAO data (see Tab. 2 of this reference) found  $f_{\rm dcdm}<0.0262$ at 95 \% C.L. and $\tau/f_{\rm dcdm}>268.8$ Gyr. Our constraints are stronger on $f_{\rm dcdm}$, compatible with the fact that we include more data, but weaker on $\tau$, which may be explained by the fact that their posteriors never quite reach $f_{\rm dcdm}\sim 1$, as necessary to derive constraints in the `very long-lived' regime.
\end{enumerate}

\begin{figure*}
    \centering
        \includegraphics[width=1.31\columnwidth]{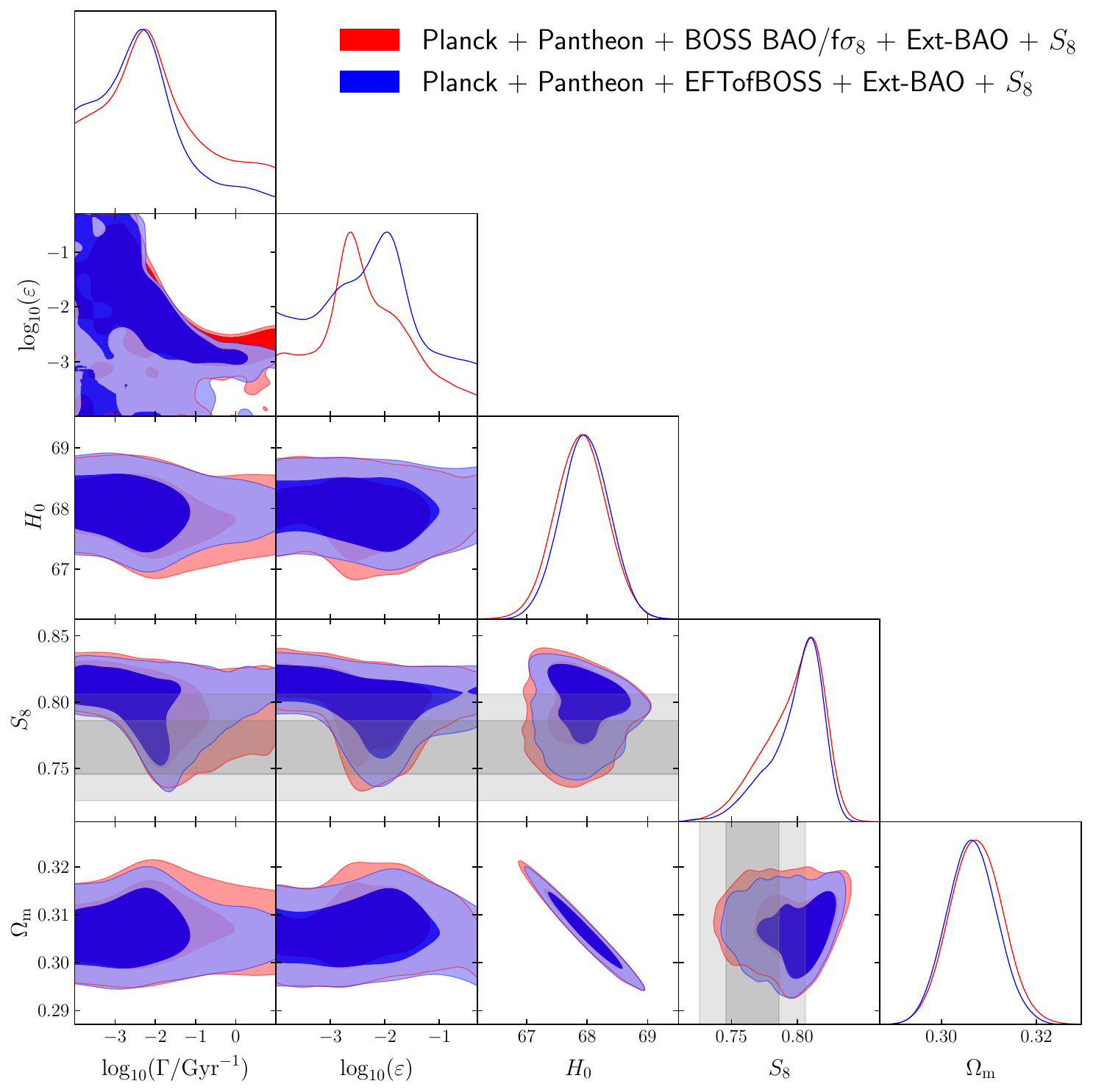}
    \caption{2D posterior distributions of the DCDM $\to$ WDM+DR model with and without the EFTofBOSS data set. We took into account the $S_8$ prior from KIDS-1000 for these two MCMC analyses. The gray shaded bands refer to the joint $S_8$ measurement from KiDS-1000 + BOSS + 2dFLens. }
    \label{fig:MCMCWDM1}
\end{figure*}

\subsection{Warm Dark Matter decay products}

\begin{table*}[t]
\begin{tabular}{|l|c|c|}
 \hline

\multicolumn{3}{|c|}{DCDM$\to$WDM+DR} \\ \hline
 Parameter & w/ EFTofBOSS  & w/ EFTofBOSS + $S_8$ \\
 \hline
 $\text{log}_{10} (\Gamma/[{\rm Gyr}^{-1}])$		     &~~~~~~~~~~unconstrained (-2.98)~~~~~~~~~~ &  ~~~~~~~~~~  $2.21(-2.08)_{-0.6}^{+1.5}$~~~~~~~~~~  \\
$\text{log}_{10} (\varepsilon)$					     & ~~~~~~~~~~unconstrained (-3.84) ~~~~~~~~~~     & $-2.30(-1.92)_{-1.10}^{+0.84}$  \\
\hline
$100 \ \omega_{\rm b} $                                    & $2.242(2.245)_{-0.014}^{+0.014}$      & $2.245(2.242)_{-0.015}^{+0.014}$ \\
$\omega_{\rm dcdm}^{\rm ini}$                        & $0.1192(0.1190)_{-0.0009}^{+0.00089}$      & $0.1188(0.1192)_{-0.00099}^{+0.00084}$ \\
$H_0 /[{\rm km/s/Mpc}]$ & $67.78(67.82)_{-0.42}^{+0.41}$ & $67.97(67.73)_{-0.42}^{+0.44}$ \\
$\text{ln}(10^{10} A_s)$                             & $3.049(3.051)_{-0.016}^{+0.015}$  & $3.046(3.052)_{-0.016}^{+0.015}$ \\
$n_s$                                                & $0.9668(0.9679)\pm0.0039$     & $0.9676(0.9670)\pm0.0039$ \\
$\tau_{\rm reio}$								     & $0.0571(0.0584)_{-0.0080}^{+0.0071}$      & $0.0564(0.0584)_{-0.0077}^{+0.0074}$ \\
\hline
$\Omega_{\rm m}$								     & $0.3090(0.3089)_{-0.0057}^{+0.0055}$     & $0.3064(0.3094)_{-0.0058}^{+0.0055}$ \\
$\sigma_8$ & $0.806(0.811)_{-0.014}^{+0.012}$ & $0.790(0.763)_{-0.010}^{+0.027}$\\
$S_8$ &	$0.818(0.823)_{-0.012}^{+0.016}$  &	        $0.798(0.775)_{-0.012}^{+0.025}$    \\
\hline
$\chi^2_{\rm min}$ & 3927.0   & 	3929.3		\\
\hline
$\footnotesize{\chi^2_{\rm min} (\text{DCDM}) -\chi^2_{\rm min} (\Lambda\text{CDM})} $ & 0.0 & $-3.8$ \\
\hline
$Q_{\rm DMAP}\equiv\sqrt{\chi^2_{\rm min}({\rm w/}~S_8)-\chi^2_{\rm min}({\rm w/o}~S_8)} $ & \multicolumn{2}{|c|}{1.5$\sigma$}\\
\hline
\end{tabular}
\caption{The mean (best-fit) $\pm 1\sigma$ errors of the cosmological parameters from our `{\it Planck} + Pantheon + EFTofBOSS + Ext-BAO' and `{\it Planck} + Pantheon + EFTofBOSS + Ext-BAO + $S_8$' analyses for the DCDM $\to$ WDM+DR model. For each data set we also report its best-fit $\chi^2$, and the $\Delta\chi^2$ with respect to the analogous $\Lambda$CDM best-fit model.}
\label{tab:WDM}
\end{table*}

\begin{table*}
\addtolength{\tabcolsep}{-1pt}
\begin{tabular}{|c|c|c|c|c|c|c|}
 \hline
 & \multicolumn{2}{|c|}{$\Lambda$CDM} &   \multicolumn{2}{|c|}{DCDM $\to$ DR} & \multicolumn{2}{|c|}{DCDM $\to$ WDM+DR} \\
\hline
 & w/ EFTofBOSS & w/ EFTofBOSS + $S_8$ & w/ EFTofBOSS & w/ EFTofBOSS + $S_8$& w/ EFTofBOSS & w/ EFTofBOSS + $S_8$\\
\hline
$\chi^2_{\text{CMASS NGC}}$ & 40.3 & 39.2 & 40.4 & 39.2 & 40.2 & 40.8\\
$\chi^2_{\text{CMASS SGC}}$ & 44.0 & 44.3 & 44.0 & 44.3 & 44.1 & 43.8\\
$\chi^2_{\text{LOWZ NGC}}$ & 33.5 & 33.5 & 33.5 & 33.5 & 33.5 & 33.7\\
\hline
$\chi^2_{\text{EFTofBOSS}}$ & 117.8 & 117.0 & 117.9 & 117.0 & 117.8  & 118.3 \\
\hline
p-value &  0.54 & 0.56 & 0.49 & 0.51  & 0.49 & 0.47  \\
\hline
\end{tabular}
\caption{$\chi^2$ of each sky-cut of the EFTofBOSS data set for our `{\it Planck} + Pantheon + EFTofBOSS + Ext-BAO' and `{\it Planck} + Pantheon + EFTofBOSS + Ext-BAO + $S_8$' analyses for $\Lambda$CDM, DCDM $\to$ DR and DCDM $\to$ WDM+DR models.}
\label{tab:chi2_EFT}
\end{table*}

We now turn to the case of the DCDM $\to$ WDM+DR model, described by the parameters $\Gamma =\tau^{-1}$, the decay rate of DCDM, and $\varepsilon$, the fraction of DCDM rest mass energy converted into DR. 
Note that in this section, we trade the density of DM today, $\omega_{\rm cdm}$, for the initial density of DM (before decays occur) at $a\to 0$, $\omega_{\text{dcdm}}^{\rm ini}$. For a stable particle, we simply have $\omega_{\text{dcdm}}^{\rm ini}\equiv\omega_{\rm cdm}$ as defined previously.
In the MCMC analyses, we imposed logarithmic priors\footnote{For discussions about the impact of prior choices, see the appendix of Ref.~\cite{abellan_2021}} on $\varepsilon$ and $\Gamma$, and a flat prior on $\omega_{\text{dcdm}}^{\rm ini}$:
\begin{align*}
    -4\le &~\text{log}_{10}(\Gamma /[{\rm Gyr}^{-1}]) \le 1, \\
    -4\le &~\text{log}_{10}(\varepsilon) \le \text{log}_{10}(0.5), \\
    0\le &~\omega_{\text{dcdm}}^{\rm ini} \le1.
\end{align*}
We present our results for the analyses with and without $S_8$ prior in Tab. \ref{tab:WDM}, while the $\chi^2$ of the EFTofBOSS data of these analysis are reported in Tab. \ref{tab:chi2_EFT}. All relevant $\chi^2$ per experiment are given in App.~\ref{sec:app_chi^2}.
In Fig. \ref{fig:MCMCWDM1}, we display the 1D and 2D posteriors of $\left\{\text{log}_{10} (\Gamma/[{\rm Gyr}^{-1}]),\text{log}_{10} (\varepsilon),H_0,S_8,\Omega_m\right\}$ for the DCDM $\to$ WDM+DR model with and without the EFTofBOSS data set, always including the $S_8$ prior. 
Posteriors without the $S_8$ prior are shown in App.~\ref{sec:app_S8}. 

\subsubsection{Estimating the tension with the $S_8$ measurement}
 Without the $S_8$ prior, the total $\chi^2$ does not show any improvement (see Tab. \ref{tab:WDM}) and the data do not favor the DCDM $\to$ WDM+DR model. In fact, in the absence of the $S_8$ prior, it seems that one could derive apparently strong constraints on these models\footnote{In Ref.~\cite{abellan_2021}, it was shown through a mock data analysis that {\it Planck} data alone could not detect the best-fit model required to explain the $S_8$ tension, artificially leading to strong constraints on the DCDM model.}.  
Yet, once the $S_8$ likelihood is included, we find  $\Delta \chi^2 = -3.8$ (for 2 extra degrees of freedom) at virtually no cost in $\chi^2$ for other likelihoods (see App.~\ref{sec:app_chi^2}): the inclusion of the $S_8$ prior help in opening up the degeneracy with the DCDM parameters, without degrading the fit to the host of cosmological data, as stressed in Refs.~\cite{Abellan:2020pmw,abellan_2021}. 
 
Nevertheless, the DCDM model is not statistically favored over $\Lambda$CDM, as the preference over $\Lambda$CDM is currently solely driven by the low $S_8$ prior, for which we have used a value only in mild $\sim 2.4\sigma$ tension with the $\Lambda$CDM prediction\footnote{Different $S_8$ priors would lead to different preferences. The preference could also be made stronger at fixed $\varepsilon$  (see Ref.~\cite{abellan_2021}).}. 
We can estimate the residual tension between data sets within the various models by computing the `difference in maximum a posterior' ($Q_{\rm DMAP}$ statistics \cite{Raveri:2018wln}) between the $\chi^2$ obtained with and without the $S_8$ prior. 
The tension estimator\footnote{In general, $Q_{\rm DMAP}$ is computed as the difference of effective $\chi^2=-2{\rm Log}{\cal L}(\theta^{\rm MAP})$, where ${\cal L}(\theta^{\rm MAP})$ is the likelihood evaluated on the maximum a posteriori $\theta^{\rm MAP}$, between the $\chi^2$ obtained in the combined analysis and the sum of the $\chi^2$ obtained in the individual analyses. For Gaussian ${\cal L}$, it is distributed as a $\chi^2$ distribution with $N_{1}+N_{2}-N_{12}$ degrees of freedom (d.o.f.), where $N_{i}$ refers to the number of d.o.f. in the individual $(i=1,2)$ and combined analysis $(i=12)$. In the case of the combination of {\it Planck} and a Gaussian prior on $S_8$, it follows a $\chi^2$-distribution with one d.o.f., and the tension can be evaluated as $Q_{\rm DMAP}\equiv\sqrt{\chi^2_{\rm min}({\rm w/}~S_8)-\chi^2_{\rm min}({\rm w/o}~S_8)}$.}
at their MAP point gives $Q_{\rm DMAP} = 1.5 \sigma$ in the DCDM$\to$WDM+DR model, as compared to $2.4\sigma$ in the $\Lambda$CDM and DCDM$\to$DR models.

\subsubsection{Impact of EFTofBOSS data}
Comparing to results without the EFTofBOSS data, for which we get\footnote{This number is different to that quoted in Refs.~\cite{Abellan:2020pmw,abellan_2021} because we recall that we make use of a different $S_8$ prior from KiDS-1000 alone, that does not include information from BOSS data and therefore has larger error bars.} $\Delta \chi^2 = -4.4$, we find that the  $\Delta \chi^2$ is only mildly degraded by the inclusion of EFTofBOSS data.
 More precisely, the $\chi^2$ of the total EFTofBOSS data for the DCDM $\to$ WDM+DR model, given in Tab. \ref{tab:chi2_EFT}, is only slightly larger than that for $\Lambda$CDM ($\Delta\chi^2=1.3$) despite a much lower $S_8\simeq0.775$ (at the best-fit) which yields a very good fit of the KiDS-1000 prior. Comparing to the analysis with the BAO/$f\sigma_8$ measurement from BOSS-DR12 (also presented in App.~\ref{sec:app_chi^2}), we note that these `compressed' data already showed a minor degradation of $\chi^2$ compared to $\Lambda$CDM ($\Delta\chi^2=1.1$). 
 We conclude that BOSS-DR12 data are in good agreement with the DCDM $\to$ WDM+DR model, but have a non-negligible impact, as the `naive' analysis presented in sec.~\ref{sec:Preliminary_study_wdm} suggested.  
 
 More precisely, one can see in Fig. \ref{fig:MCMCWDM1} that the main impact of EFTofBOSS data is to cut in the $\log_{10}(\Gamma/{\rm  Gyr}^{-1})-\log_{10}(\varepsilon)$ degeneracy, excluding too large values of $\log_{10}(\Gamma/{\rm Gyr}^{-1})$. In App. \ref{sec:app_WDM_LCDM} we show that including the EFTofBOSS data does not shift the $\Lambda$CDM parameters. 
Therefore, the EFTofLSS significantly improves the constraints on the $\tau= \Gamma^{-1}$ parameter at $1\sigma$: 
\begin{align}
   &1.61 < \rm{log}_{10}(\tau /{\rm Gyr}) < 3.71~~~~(w/ {\rm EFTofBOSS})\,, \nonumber
\end{align}
to be compared with
\begin{align}
   &1.31 < \rm{log}_{10}(\tau /{\rm Gyr}) < 3.82~~~(w/o~{\rm EFTofBOSS} )\,.\nonumber
\end{align}

 Additionally, we observe a notable evolution of the DCDM parameters of the best-fit model compared to the analysis without EFTofBOSS (and with the BAO/$f\sigma_8$ measurement from BOSS-DR12 instead): the best-fit model, with the inclusion of the $S_8$ likelihood, now has $\Gamma = 0.0083 \ \rm{Gyr}^{-1}$ ($\tau = 120 $ Gyr) and $\varepsilon = 0.012 $, while previously $\Gamma = 0.023 \ \rm{Gyr}^{-1}$ ($\tau = 43$ Gyr) and $\varepsilon = 0.006$.
 This means that EFTofBOSS data favors longer lived DM models and therefore a smaller fraction of WDM today $f_{\rm wdm}\equiv \bar{\rho}_{\rm wdm}/(\bar{\rho}_{\rm dcdm}+\bar{\rho}_{\rm wdm})\simeq 10 \%$ compared to $f_{\rm wdm}\simeq 27 \%$ previously, but a significantly larger kick velocity $v_{\rm kick}/c \simeq \varepsilon$ (and therefore a larger free-streaming scale).

It is instructive to compare these numbers with recent constraints derived from observations of Milky Way satellites by the DES collaboration \cite{Mau:2022sbf}. These constraints exclude ${\rm log}_{10}(\Gamma/{\rm Gyr}^{-1})\gtrsim -1.5$ for ${\rm log}_{10}(v_{\rm kick}/c) \simeq {\rm log}_{10}(\varepsilon) \gtrsim -4$. The best-fit model of our EFTofBOSS analysis, and a large fraction of the 68\% C.L., lie well within the allowed region, but these observations certainly provide a crucial test of the DCDM cosmology, as a deficit of satellites compared to $\Lambda$CDM is expected in this model.

\begin{figure*}
    \centering
\includegraphics[trim=1cm 0cm 3cm 1cm, clip=true, width=2\columnwidth]{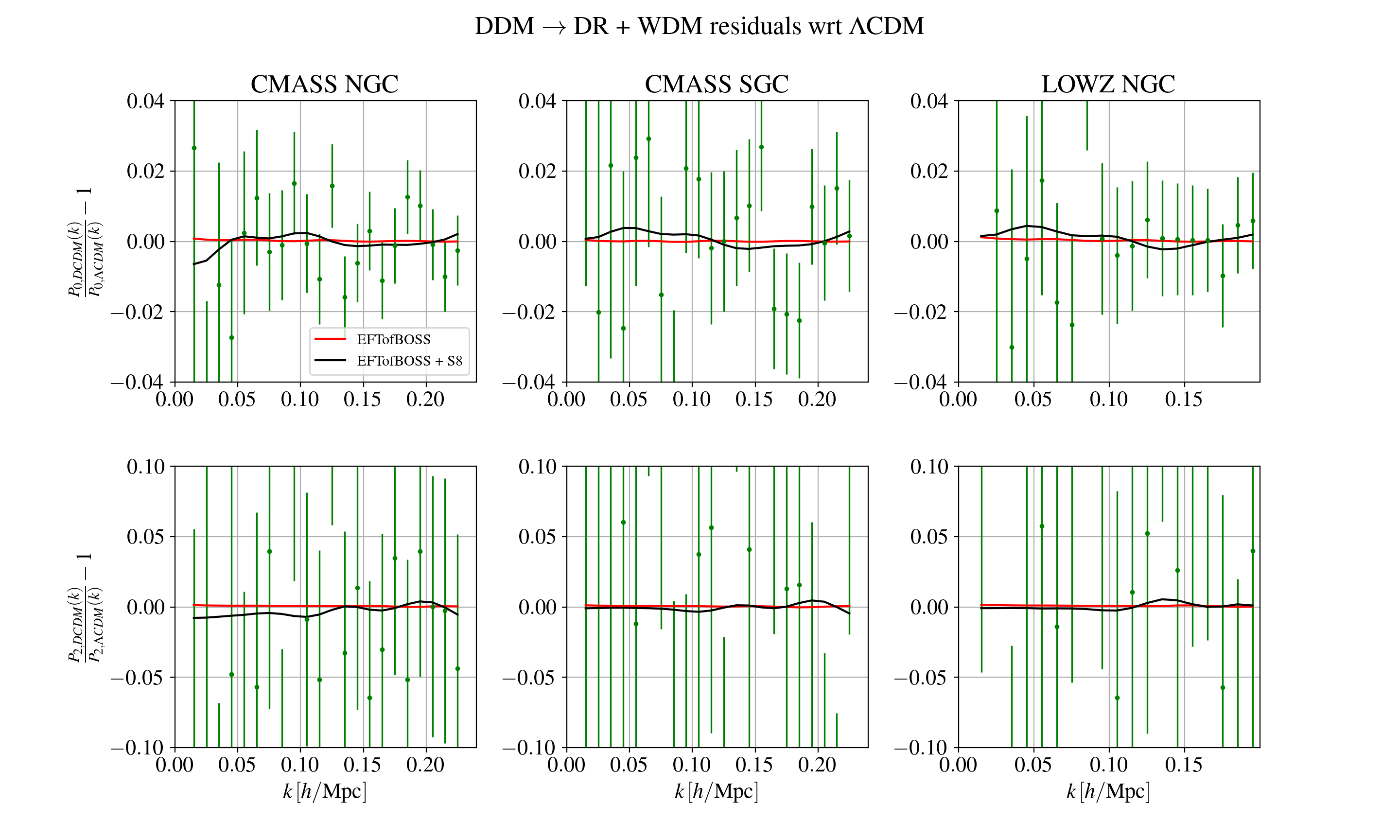}
\caption{Residuals of the monopole and the quadrupole of the DCDM $\to$ WDM+DR model for EFTofBOSS data and EFTofBOSS data + $S_8$ prior. We normalized these residuals as well as the data with the $\Lambda$CDM best-fit (EFTofBOSS data).}
    \label{fig:WDM_residuals}
\end{figure*}

\begin{figure}
    \centering
    \includegraphics[trim=1.5cm 0cm 19cm 11cm, clip=true, width=1\columnwidth]{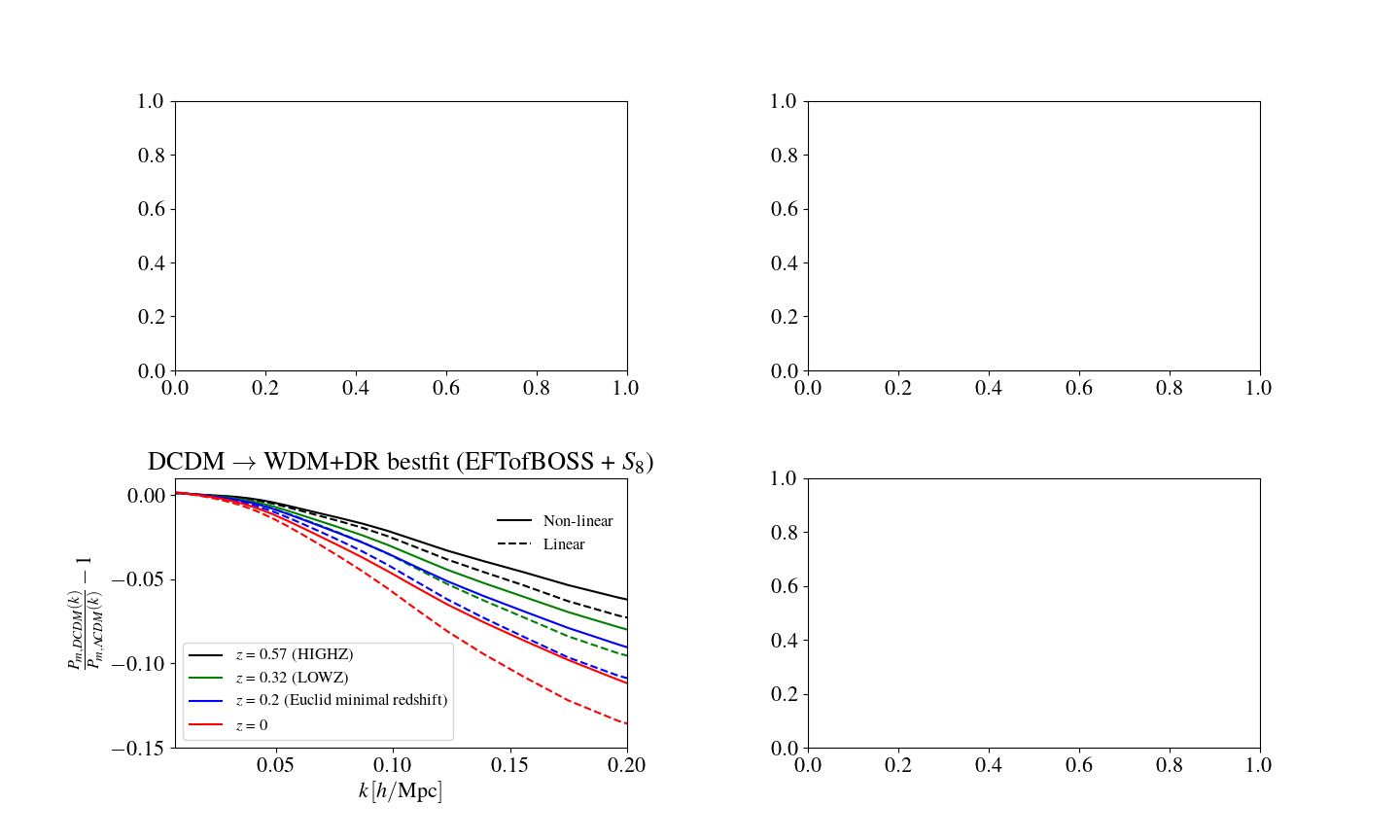}
    \caption{Residuals of linear (dashed lines) and non-linear (solid line) matter power spectrum of the DCDM $\to$ WDM+DR model (EFTofBOSS data + $S_8$ prior) for $z=$ 0, 0.2 (Euclid minimal redshift), 0.32 (effective redshift of the LOWZ sky-cut) and 0.57 (effective redshift of the HIGHZ sky-cut). We normalized these residuals with the $\Lambda$CDM best-fit (EFTofBOSS data).}
    \label{fig:WDM_different_z}
\end{figure}

\subsubsection{Towards high-accuracy measurements of the galaxy power spectrum}

To gauge the importance of future surveys in constraining the  DCDM $\to$ WDM+DR model, we show in Fig. \ref{fig:WDM_residuals} the residuals of the monopoles and quadrupoles of the galaxy power spectrum between the  DCDM $\to$ WDM+DR and $\Lambda$CDM models. 
One can see that there are sub-percent differences between the models that gives us hope to probe the DCDM model further. 
Indeed, future galaxy clustering power spectrum data with higher precision and measurements at additional redshift bins such as Euclid \cite{Amendola:2016saw}, VRO \cite{Mandelbaum:2018ouv} and DESI \cite{Aghamousa:2016zmz} have an expected sensitivity that should allow us to detect these mild differences. 
In order to estimate the  impact of future observations on the preference of the DCDM $\to$ WDM+DR model with respect to the $\Lambda$CDM model, we plot in Fig.~\ref{fig:WDM_different_z} the residuals of the non-linear matter power spectrum\footnote{We set here $c_{s} = 1$, which is an effective parameter of the one-loop correction that can be interpreted as the effective sound speed of the dark matter.} between the best-fit of the DCDM $\to$ WDM+DR model (for the `EFTofBOSS + $S_8$' analysis) and $\Lambda$CDM model (for the `EFTofBOSS' analysis). 
We represent it for different redshifts, starting at the minimal redshift probed by an experiment like Euclid \cite{Amendola:2016saw}. 
Note that at the level of the non-linear matter power spectrum, the suppression with respect to the $\Lambda$CDM model at $z=0.32$ and $z=0.57$ corresponding to current observations are more than one order of magnitude stronger than what is seen in the residual of the monopole and quadrupole of the galaxy power spectrum (see Fig. \ref{fig:WDM_residuals}). This is due to the impact of the degeneracy between the DCDM parameters and the EFT galaxy bias parameters, which can counteract the effect of the DCDM decay in the galaxy power spectrum.
This shows that current theoretical uncertainties associated with galaxy bias parameters limit the ability to use galaxy (clustering) surveys to probe the DCDM model, and represent a potential challenge to fully exploit future surveys. 
Additionally, we observe that as $z$ decreases, the deviation from $\Lambda$CDM increases significantly because of the production of WDM through the decay. 
We keep for future work to check through dedicated forecasts whether accumulation of low redshift data, as well as the reduction of error bars, will allow us to firmly detect or exclude the DCDM $\to$ WDM+DR model that resolves the $S_8$ tension.

\section{\label{sec:conclusion}Conclusions}
In this paper, we have confronted two models of DCDM with BOSS DR12 galaxy power spectrum data \cite{Alam:2016hwk} as described by the EFTofLSS from Refs.~\cite{senatore_2015, Mirbabayi_2015, Angulo_2015, Fujita_2020, perko_2016,Nadler_2018, D_Amico_2021}. 
We focused first on a model where a fraction of dark matter decays into dark radiation, the DCDM $\to$ DR model, and second on a model where all the dark matter decays into warm massive particles and dark radiation particles, the DCDM $\to$ WDM+DR model. 
The latter model was recently suggested as a possible resolution to the $S_8$ tension, the mismatch between the determination of the $S_8$ parameter from {\it Planck} CMB power spectrum \cite{Aghanim:2018eyx} and from weak lensing surveys by KiDS \cite{Heymans:2020gsg,KiDS:2020suj}, CFHTLenS \cite{Heymans:2013fya} and DES \cite{DES:2021wwk}.  
We presented in Sec.~\ref{sec:NL} the first calculation of the non-linear (matter and galaxy) power spectra in DCDM models making use of recent progresses in the EFTofLSS. 
We then confronted in Sec.~\ref{sec:DataAnalysis} these two models to a compilation of {\it Planck} TTTEEE and lensing power spectra,  BAO data from BOSS and eBOSS (including Ly-$\alpha$ data), uncalibrated luminosity distance to SN1a from the Pantheon catalogue \cite{Scolnic:2017caz}, as well as measurements of the monopole and quadrupole of the galaxy power spectrum for three different sky-cut of BOSS-DR12 (see Ref. \cite{zhang2021boss}), namely  LOWZ NGC, CMASS NGC and CMASS SGC \cite{Alam:2016hwk}. 
We compared the use of either the BAO/$f\sigma_8$ from that same release, or the full shape of the galaxy power spectrum. Additionally, we tested the ability of these models to resolve the $S_8$ tension by performing analyses with and without prior on $S_8$ as measured by KiDS \cite{KiDS:2020suj}. 
Our results can be summarized as follows:

\begin{enumerate}
    \item[\textbullet]  We have derived the most up-to-date bound on the fraction of decaying dark matter $f_{\rm dcdm}$, which is now $f_{\rm dcdm}<0.0216$ for short-lived DCDM. 
    We have also updated constraints on the lifetime of dark matter for the case where $f_{\rm dcdm}\to1$, namely $\tau/f_{\rm dcdm}>249.6$ Gyr.  
    However, we have found that the EFTofLSS does not provide significantly better constraints to the cosmological parameters for the DCDM $\to$ DR model, compared to the use of the standard BAO/$f\sigma_8$ data.
    In agreement with past studies, we have found that these models do not help neither for the $S_8$ nor for the $H_0$ tension, and the inclusion of EFTofBOSS data doesn't alter that conclusion.
    \item[\textbullet] The DCDM $\to$ WDM+DR model can explain the low $S_8$ value measured by KiDS-1000 while preserving the goodness of fit to other data set, including EFTofBOSS data. The residual tension is $1.5\sigma$ compared to  $2.4\sigma$ in the $\Lambda$CDM model.
    Nevertheless, the model is not statistically favored over $\Lambda$CDM  ($\Delta\chi^2=-3.8$ for 2 degrees of freedom, roughly corresponding to $1.5\sigma$).  
    The inclusion of EFTofBOSS data only marginally affects the preference.
    \item[\textbullet] EFTofBOSS data however do significantly improve the 1-$\sigma$ constraint on the DCDM lifetime for the DCDM $\to$ WDM+DR model, and when combined with the $S_8$ prior, we now obtain $\rm{log}_{10}(\tau / \rm{Gyr}) = 2.21_{-0.6}^{+1.5}$  compared to $\rm{log}_{10}(\tau / \rm{Gyr}) =1.92_{-0.61}^{+1.9}$ without the EFTofBOSS.  The constraints on $\rm{log}_{10}(\varepsilon)$ are however slightly weaker than with BAO/$f\sigma_8$ measurements.
    \item[\textbullet] The EFTofBOSS data also affects the best-fit model which, with the inclusion of the $S_8$ likelihood, corresponds to a longer lived DM with $\tau = 120$ Gyr (compared to $\tau = 43$ Gyr previously) and a larger kick velocity $v_{\rm kick}/c \simeq \varepsilon = 1.2\% $ (compared to $v_{\rm kick}/c  \simeq 0.6\% $ previously). 
\end{enumerate}

Looking forward, we expect future galaxy clustering power spectrum data, with higher precision and measurements at additional redshift bins such as Euclid \cite{Amendola:2016saw}, VRO \cite{Mandelbaum:2018ouv} and DESI \cite{Aghamousa:2016zmz}, to provide us with exquisite sensitivity to DM decays into an invisible sector whether massive or massless. Moreover, as the error bars decrease, it will likely be necessary to identify and account for the corrections to be made to the EFTofLSS in order to capture all the specific effects of the DCDM $\to$ WDM+DR model.
Following Ref. \cite{senatore2017neutrinos} for the case of massive neutrinos, it will be important to determine the one-loop terms and associated counterterms of the mildly non-linear galaxy power spectrum caused by the
WDM contribution to the linear matter power spectrum (which we have argued in App. \ref{sec:app_EFT_WDM} to likely be small compared to current error bars). We keep for future work to test whether these surveys will be able to firmly detect or exclude the DCDM $\to$ WDM+DR model that resolves the $S_8$ tension. 

\begin{acknowledgements}

The authors are thankful to Pierre Zhang for his precious help with the PyBird code. His explanations and advices were essential to the success of this project. We also warmly thank Zakaria Chacko and Abhish Dev for their contribution in the early stages of this project.
This work has been partly supported by the CNRS-IN2P3 grant Dark21. 
The authors acknowledge the use of computational resources from the Excellence Initiative of Aix-Marseille University (A*MIDEX) of the “Investissements d’Avenir” programme. 
This project has received support from the European Union’s Horizon 2020 research and innovation program under the Marie Skodowska-Curie grant agreement No 860881-HIDDeN. 
P.D. is supported in part by Simons Investigator in Physics Award 623940 and NSF award PHY-1915093. 
Y.T. is supported by the NSF grant PHY-2014165 and PHY-2112540.

\end{acknowledgements}

\appendix

\section{Comparison between the EFTofLSS and N-body methods for the DCDM $\to$ DR model}
\label{sec:app_EFT_vs_Nbody}

\begin{figure*}
    \centering
\includegraphics[trim=2cm 0cm 3cm 11cm, clip=true, width=2\columnwidth]{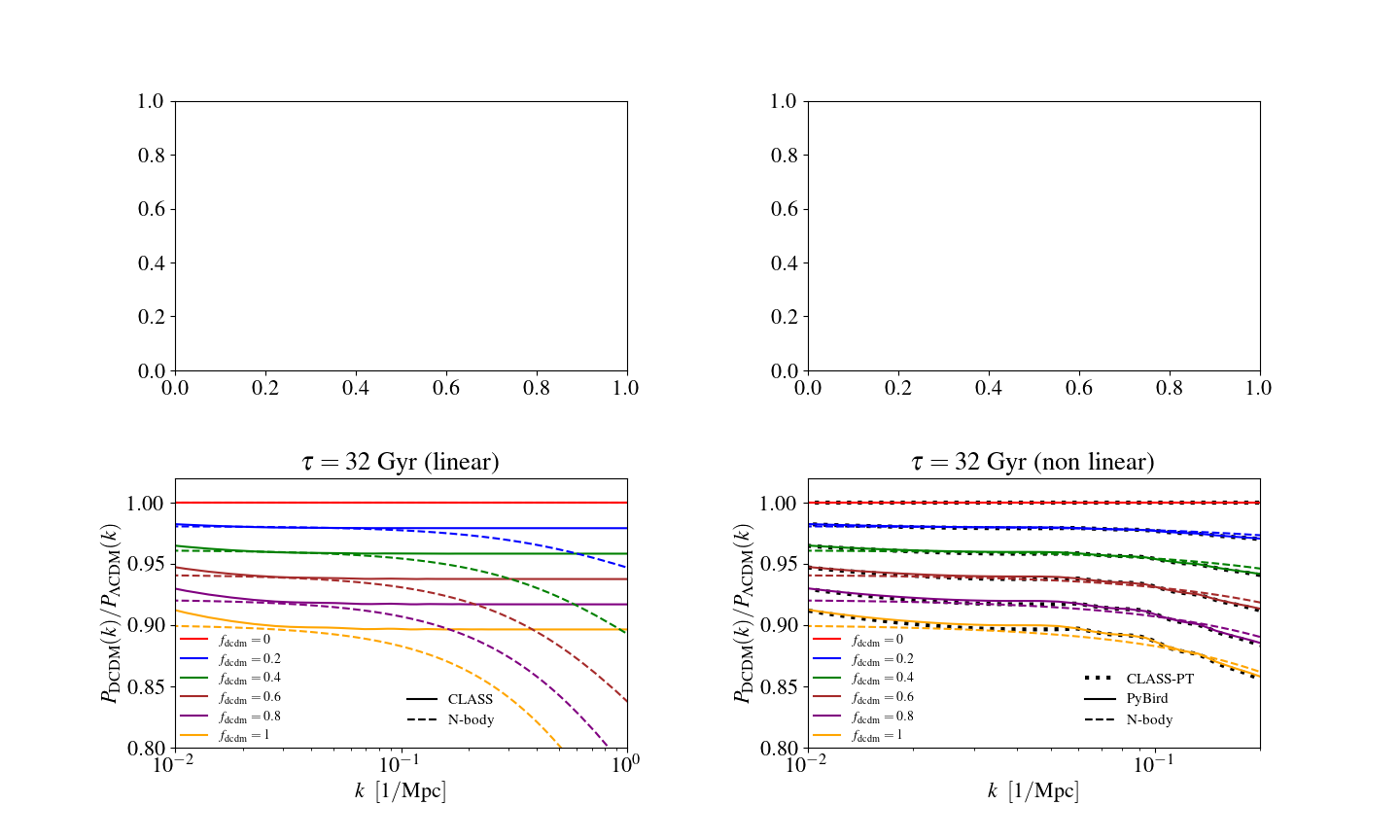}
\caption{Comparison between the residuals of the non-linear matter power spectra predicted by the N-body simulation and the residuals of the linear matter power spectra predicted by the CLASS code on the one hand (left panel), and the residuals of the non-linear matter power spectra predicted by the CLASS-PT and PyBird codes on the other hand (right panel). We compute these power spectra for $z=0$ and for $\tau = 32$ Gyr, while we varied $f_{\rm dcdm}$ from 0 to 1 with a step of 0.2.}
    \label{fig:Nbody_vs_CLASSPT}
\includegraphics[trim=1.5cm 0cm 19cm 11cm, clip=true, width=1.15\columnwidth]{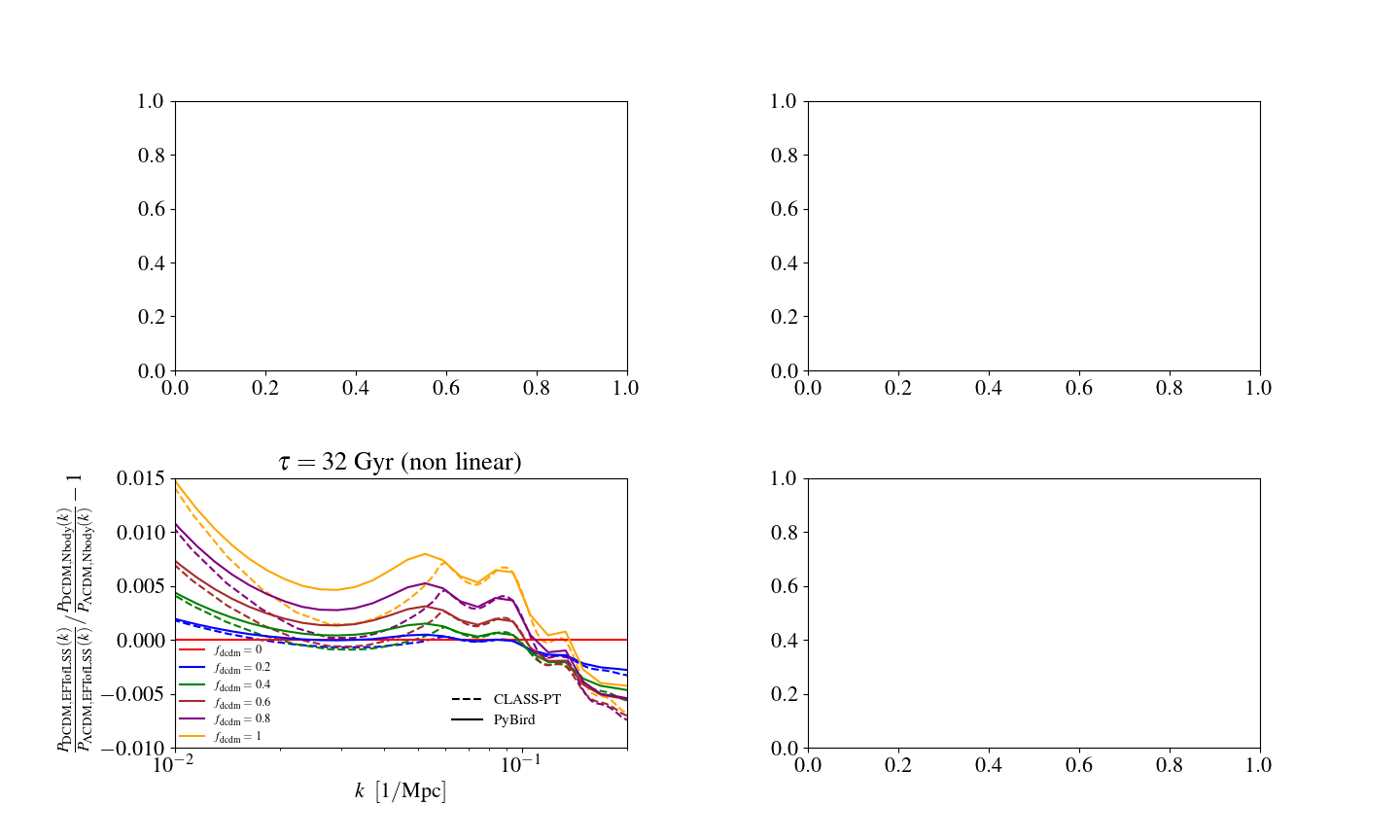}
\caption{Ratio between residuals of the non-linear matter power spectra obtained from the N-body simulation and those obtained from the CLASS-PT and the PyBird codes for $z=0$, $\tau = 32$ Gyr and $f_{\rm dcdm}$ varying from 0 to 1 with a step of 0.2.}
    \label{fig:Nbody_vs_CLASSPT_residuals}
\end{figure*}

In this appendix,  we compare the non-linear matter power spectrum obtained through the EFTofLSS method with the results of dedicated N-body simulations performed in Ref. \cite{hubert2021decaying}. 
Authors of Ref. \cite{hubert2021decaying} have determined a fitting formula which describe the correction to the non-linear matter power spectrum due to the DM decay compared to the $\Lambda$CDM model, as a function of $\tau$, $f_{\rm dcdm}$, and the redshift $z$.
In Fig. \ref{fig:Nbody_vs_CLASSPT}, we compare this fitting formula, where we set $\tau = 32$ Gyr and $z=0$ and vary $f_{\rm dcdm} \in [0,1]$, with the linear matter power spectrum of the CLASS code (left panel), and with the non-linear matter power spectra from both the CLASS-PT and PyBird codes (right panel). 
Here, we set the $\Lambda$CDM parameters to the values used in Ref. \cite{hubert2021decaying}. 
The left panel of this figure is intended as a reproduction of Fig. 1 of this reference for direct comparison, while the right panel presents the comparison of interest. 
Indeed, from the right panel of Fig. \ref{fig:Nbody_vs_CLASSPT}, one can clearly see that (i) the CLASS-PT and PyBird codes give very similar power spectra for the DCDM $\to$ DR model\footnote{We set, in the PyBird and CLASS PT codes, $c_{s} = 1$, which is an effective parameter of the one-loop correction that can be interpreted as the effective sound speed of the dark matter.}, (ii) the deviation from $\Lambda$CDM predicted in these two EFTofLSS codes is very close to that obtained through N-body simulation. 
In order to determine more precisely the deviations between the EFTofLSS and the N-body methods, we plot, in Fig. \ref{fig:Nbody_vs_CLASSPT_residuals}, the ratio between the residuals obtained with the N-Body simulation and those obtained with the CLASS-PT and PyBird codes. 
One can see that the difference is below the $\sim 1\%$ level until $k\sim0.2 \ h\,\text{Mpc}^{-1}$ (the maximum $k$ at which the EFTofLSS is valid at one-loop order for a small $z$). 
Let us note that the difference between the N-body simulation and the EFTofLSS power spectrum for $k \lesssim 0.02 \ h\,\text{Mpc}^{-1}$ is not relevant; it is merely due to the fact that the N-body fitting formula does not encode this behavior for low $k$ (see \cite{hubert2021decaying}), but this $k$-range is well within the linear regime and does not necessitate a correction. 
Let us also remark that the lower $f_{\rm dcdm}$ (or the longer $\tau$), the smaller the difference between the residuals from the N-body and those from the EFTofLSS method. 
Since current constraints only allow small values of $f_{\rm dcdm}\lesssim 2.5\%$ or large lifetime $\tau\gtrsim 240$ Gyr, it is safe to use the PyBird (or CLASS-PT) code in their current form to describe the DCDM$\to$DR model. 
This good agreement between the EFT approach and the N-body simulation, despite having made no change to the EFT modelling, may appear surprising at first sight. However, there is a fairly intuitive argument as to why the DM equations (and therefore the EFT formalism) should receive only minor corrections from the presence of a non-zero decay term.   This is because, in the synchronous gauge at linear order, the DCDM equations are strictly identical to that of CDM: the effect of the decay is happening exactly at the same rate everywhere in space, and therefore cancels out of the perturbed continuity and Euler equations which drive the DCDM perturbed dynamics.  Although strictly speaking, the contribution of the decay term may appear at higher order, as we treat the mildly non-linear regime, it will be sub-dominant. This explains why we find such a good agreement between N-body simulations and the EFT computation despite not modifying the master equations, the expansion nor the counterterms. Note that this argument is valid irrespective of the mass of the daughter particles as far as the mother particle is concerned.  Similarly, corrections to the massless daughter equations may appear, but will likely have only a small impact on the observables given that the massless daughter quickly redshift away compared to other species for decays happening at late-times (at times relevant for galaxy surveys).

\section{Assessing the validity of the EFTofLSS in the DCDM $\to$ WDM+DR model}
\label{sec:app_EFT_WDM}

\begin{figure*}
    \centering
\includegraphics[width=1.03\columnwidth]{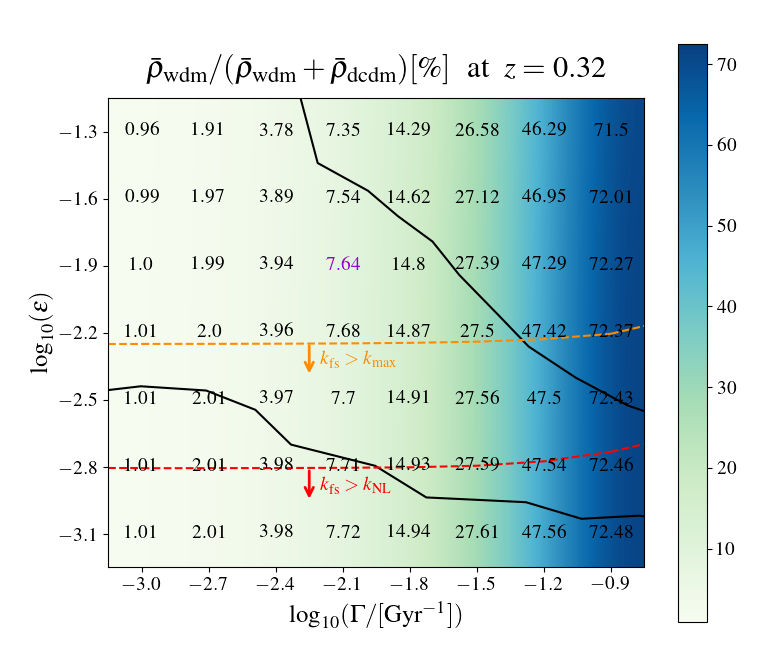}
\includegraphics[width=1.03\columnwidth]{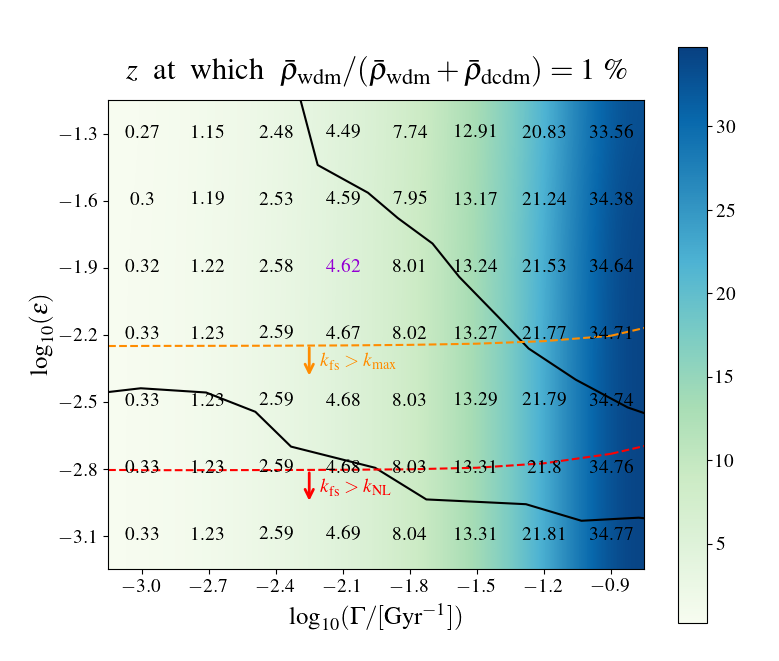}
\caption{Values of the WDM fraction at $z=0.32$ (left pannel) and the redshift at which the WDM fraction becomes 1\% (right pannel), in a region of the  $\rm{log}_{10}(\varepsilon)$-$\rm{log}_{10}(\Gamma/ \rm{Gyr}^{-1} )$ plane. The $\Lambda$CDM parameters are fixed to the best-fit from the `{\it Planck} + Pantheon + EFTofBOSS + Ext-BAO + $S_8$' analysis. The black lines indicate the $1\sigma$ limits of this analysis, while the point highlighted in purple indicates the best-fit. All the points below the yellow and red lines correspond to models having a WDM free-streaming wavenumber $k_{\rm fs}$ larger than  $k_{\rm max}=0.2 \ h\,\text{Mpc}^{-1}$ and $k_{\rm nl}=0.7 \ h\,\text{Mpc}^{-1}$, respectively.}
    \label{fig:EFT_WDM_validity}
\end{figure*}

In this appendix, we discuss the validity of the EFTofLSS in the DCDM $\to$ WDM+DR model. 
In Ref.~\cite{senatore2017neutrinos}, the EFTofLSS was extended to describe massive neutrinos, an extension to $\Lambda$CDM with properties similar to that of the DCDM$\to$WDM+DR model. 
Indeed, the main issue with employing the EFTofLSS to describe the DCDM$\to$WDM+DR model does not lie in the effect of the decay itself (the effect of the decay on the perturbed equations of the DCDM is identical to that of the DCDM $\to$ DR model, which is captured by our formalism as discussed in App. \ref{sec:app_EFT_vs_Nbody}), 
but rather in the production of a warm massive species which may contribute in a non-trivial way to the power spectrum of galaxies.
At the linear level, it was found the massive decay products behave similarly to CDM at wavenumbers smaller than the free-streaming scale $k_{\rm fs}$ (with $k_{\rm fs}$ approximately given by Eq.~\ref{eq:kfs}), but is strongly suppressed due to pressure terms at larger wavenumbers similarly to WDM and hot DM such as neutrinos.
In Ref.~\cite{senatore2017neutrinos}, the contribution of neutrinos to the total one loop power spectrum was computed, and it was found that the dominant effect is a correction to the dark matter power spectrum that scales like $16f_\nu$, where $f_\nu \equiv \bar{\rho}_\nu/(\bar{\rho}_\nu+\bar{\rho}_{\rm cdm})\sim 1 \%$, at $k>k_{\rm fs}$ and roughly half of that at $k<k_{\rm fs}$. 
The naive ${\cal O}(f_\nu)$ contribution is enhanced by twice the logarithm of the redshift of matter-radiation equality, as neutrinos are present from early times. 
The log-enhanced contribution represents about $70\%$ of the contribution to the total one loop power spectrum. 
Additionally, at leading order, counter-terms can be captured by simply re-scaling the effective DM sound speed $c_s^2$ and do not necessitate adding new parameters to the dark-matter-only calculation.
In the case of the DCDM model, the WDM is produced at much later times. 
We plot in Fig.~\ref{fig:EFT_WDM_validity} (right panel) the redshift $z_{1\%}$ at which the WDM contribution $f_{\rm wdm}\equiv \bar{\rho}_{\rm wdm}/(\bar{\rho}_{\rm wdm}+\bar{\rho}_{\rm dcdm})$  reaches $\sim 1\%$. 
We also represent the limit at 68\% C.L. derived in our work. 
For the best-fit model (shown in purple in the figure), $z_{1\%}\sim5$.
The log-enhancement from the ratio of scale factor between $z\sim5$ and $z_{eff,{\rm LOWZ}}=0.32$ is $\log[(1+z_{1\%})/(1+z_0)]\approx 1-2$ compared to the $\log[(1+z_{\rm eq})/(1+z_0)]\approx 8$ in the neutrino study in Ref.~\cite{senatore2017neutrinos} that gives the $16f_\nu$ factor. 
We therefore expect the WDM correction to be comparable to the massive neutrino case even if the energy density (today) ratio is $\approx10$ times larger than neutrinos.

\

We plot in Fig.~\ref{fig:EFT_WDM_validity} left panel, the fractional contribution of WDM at $z=0.32$ (the effective redshift of the low-z surveys) as a function of $\Gamma$ and $\varepsilon$.  
We also represent the limit at 68\% C.L. derived in this work. 
One can see that it is under $\sim15\%$ as long as $\textrm{log}_{10}(\Gamma/[{\rm Gyr}^{-1}])\lesssim -1.8$. 
Additionally, we show the value of $\varepsilon-\Gamma$ for which the free-streaming scale $k_{\rm fs}$ is equal to the maximum $k$ mode relevant for our analysis of BOSS data ($k_{\rm max}=0.2 \ h\,\text{Mpc}^{-1}$) and the maximum scale considered in the EFT computation ($k_{\rm nl}=0.7 \ h\,\text{Mpc}^{-1}$). 
In a large part of the parameter space favored by our analysis for which $f_{\rm wdm}$ is not small, $k_{\rm fs}$ exceeds  $k_{\rm max}$ and therefore corrections should also be minor. 
An improved EFT treatment including the effect of the massive decay product would be necessary however to describe the power spectrum up to $k_{\rm nl}$. 
Given current precision of the data and the large theoretical uncertainty already present, the corrections to our calculation should be negligible, but more work needs to be done to accurately describe the part of the parameter space with large $\Gamma$ (and leading to large $f_{\rm wdm}$), in particular for future surveys which can reach sub-percent precision at larger wave-numbers. 

\section{The role of the $S_8$ prior}

\begin{figure*}
    \centering
\includegraphics[width=1.30\columnwidth]{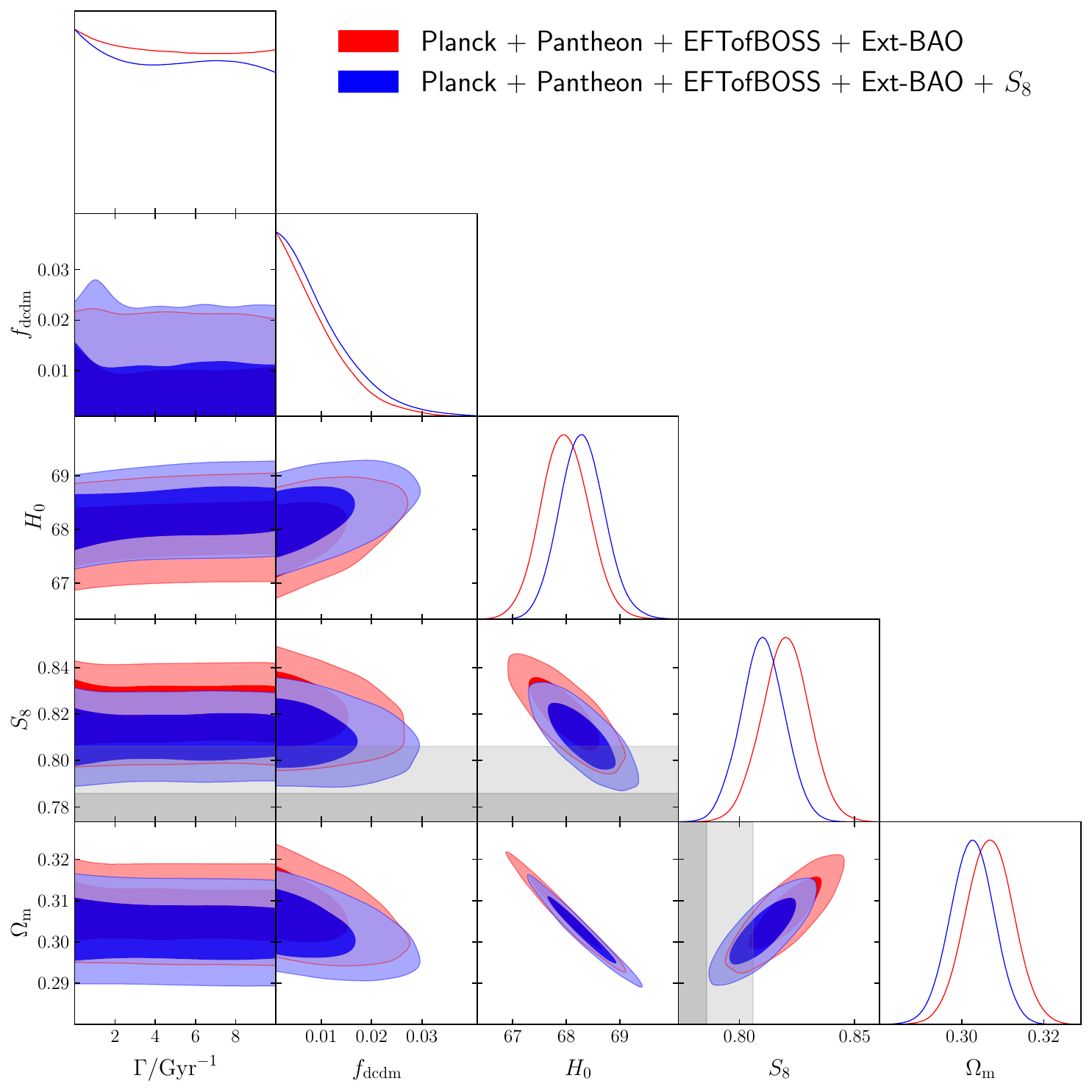}
        \caption{2D posterior distributions of the DCDM $\to$ DR model reconstructed from an analysis of {\it Planck}, Pantheon, Ext-BAO and EFTofBOSS data, with (blue) and without (red) the $S_8$ prior from KiDS-1000. The gray shaded bands refer to the joint $S_8$ measurement from KiDS-1000 + BOSS + 2dFLens.  }
    \label{fig:MCMCDR1_S8}
\end{figure*}

\begin{figure*}
    \centering
\includegraphics[width=1.31\columnwidth]{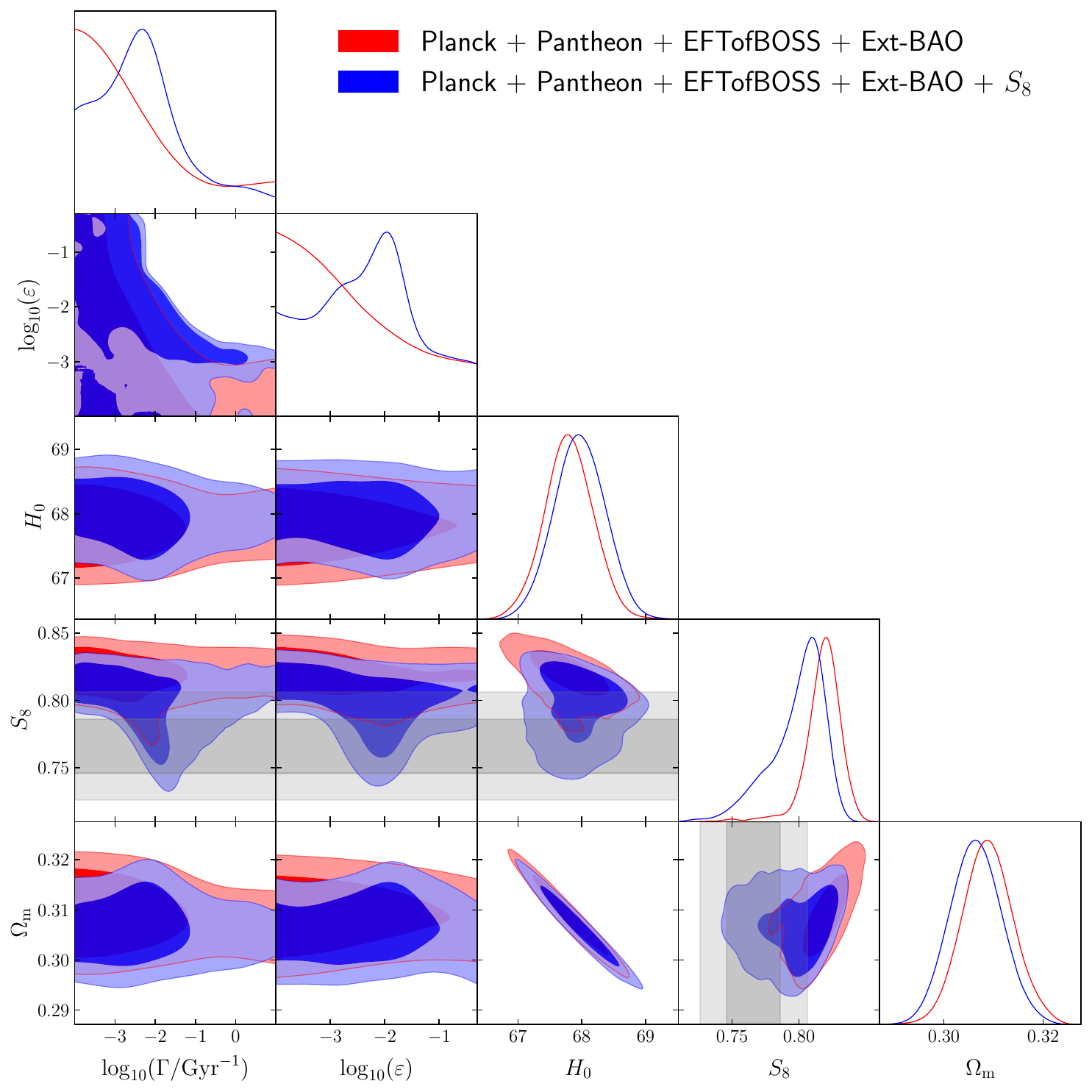}
    \caption{2D posterior distributions of the DCDM $\to$ WDM+DR model reconstructed from an analysis of {\it Planck}, Pantheon, Ext-BAO and EFTofBOSS data, with (blue) and without (red) the $S_8$ prior from KiDS-1000. The gray shaded bands refer to the joint $S_8$ measurement from KiDS-1000 + BOSS + 2dFLens.}
    \label{fig:MCMCWDM1_S8}
\end{figure*}

In this appendix, we present 2D posterior distributions obtained with and without the $S_8$ prior (but with the EFTofBOSS data) in both DCDM cosmologies. In the case of the DCDM $\to$ DR model, represented in Fig. \ref{fig:MCMCDR1_S8}, the impact of the $S_8$ prior is minor. However, in the case of the DCDM $\to$ WDM+DR model, represented in Fig. \ref{fig:MCMCWDM1_S8}, it opens up a degeneracy with $\{\Gamma,\varepsilon\}$ which can lead to low $S_8$ while preserving the fit to other data sets. Without the $S_8$ prior, the DCDM model is not favored by {\it Planck} data. 
As discussed in the main text, when the $S_8$ prior is included, the fit to {\it Planck} data is not affected, while the DCDM model can accommodate the lower $S_8$ value, contrarily to the $\Lambda$CDM model. 
From the $Q_{\rm DMAP}$ statistics \cite{Raveri:2018wln}, we can estimate the residual tension as $Q_{\rm DMAP}\equiv\sqrt{\chi^2_{\rm min}({\rm w/}~S_8)-\chi^2_{\rm min}({\rm w/o}~S_8)} = 1.5 \sigma$ in the DCDM$\to$WDM+DR model, as compared to $2.4\sigma$ in the $\Lambda$CDM model and DCDM$\to$DR model.

\label{sec:app_S8}

\section{Supplementary tables of $\chi^2_{\rm min}$ values per experiment}
\label{sec:app_chi^2}

In this appendix, we report the best-fit $\chi^2$ per experiment for both $\Lambda$CDM (Tab. \ref{tab:chi2_total_LCDM}) and DCDM $\to$ WDM+DR (Tab. \ref{tab:chi2_total_WDM}) models for our analyses with the `BAO/$f\sigma_8$ + $S_8$', `EFTofBOSS' and `EFTofBOSS + $S_8$' data.
To help the reader gauge the goodness of fit, the number of d.o.f. is estimated to be 2287 for {\it Planck} high-$l$ TTTEEE, 25 for {\it Planck} low-$l$ EE, and 25 for {\it Planck} low-$l$ TT \cite{Aghanim:2018eyx}. Other experiments do not report the number of degrees of freedom, but it can be estimated from the number of data points $N_{\rm data}$, assuming uncorrelated data-points for simplicity, and the number of free parameters $N_{\rm param} = N_{\rm param, model}+ N_{\rm param, nuisance}$, as $N_{\rm dof}=N_{\rm data}-N_{\rm param}$. In practice, we have $N_{\rm data} = 1048$ and $N_{\rm param, nuisance} = 1 $ for Pantheon \cite{Scolnic:2017caz}, $N_{\rm data} = 132$ and $N_{\rm param, nuisance} = 6$ for the sum of the 3 sky-cuts of the EFTofBOSS data including BAO, and $N_{\rm data}  = 13$ for the BOSS BAO/$f\sigma_8$ and Ext-BAO (the full BAO data set). Finally, we have for each model $N_{{\rm param},\Lambda{\rm CDM}}=6$ and $N_{{\rm param},\Lambda{\rm DCDM}}=8$.

\begin{table*}
\begin{tabular}{|l|c|c|c|}
 \hline
\multicolumn{4}{|c|}{$\Lambda$CDM} \\ \hline
 Data set & w/ BAO/$f\sigma_8$ + $S_8$ & w/ EFTofBOSS  & w/ EFTofBOSS + $S_8$ \\
\hline
{\it Planck} high-$l$ TTTEEE & 2349.0 & 2347.4 & 2349.3\\

{\it Planck} low-$l$ EE & 396.1 & 396.7 & 396.1 \\

{\it Planck} low-$l$ TT & 22.8 & 23.0 & 22.7 \\

{\it Planck} lensing & 9.6 & 8.9 & 9.6 \\

Pantheon & 1027.0 & 1027.1 & 1027.0 \\

Ext-BAO & 6.3 & 6.2 & 6.3 \\

BOSS BAO/$f\sigma_8$  & 6.0& $-$ &  $-$ \\

EFTofBOSS & $-$  & 117.8 & 117.0 \\

$S_8$ & 5.3 & $-$ & 5.0 \\

\hline
total $\chi^2_{\rm min}$ & 3821.9   & 3927.0	&	3933.0	\\
\hline

\end{tabular}
\caption{$\chi^2$ of each data set for our `{\it Planck} + Pantheon + BOSS BAO/$f\sigma_8$ + Ext-BAO + $S_8$', `{\it Planck} + Pantheon + EFTofBOSS + Ext-BAO' and `{\it Planck} + Pantheon + EFTofBOSS + Ext-BAO + $S_8$' analyses for the $\Lambda$CDM model. Since we rounded the $\chi^2$ of each experiment, the total $\chi^2$ is only equal to the sum of each $\chi^2$ at ${\cal O}(0.1)$ precision.} 
\label{tab:chi2_total_LCDM}
\end{table*}

\begin{table*}

\begin{tabular}{|l|c|c|c|}
 \hline
\multicolumn{4}{|c|}{DCDM$\to$WDM+DR} \\ \hline
 Data set & w/ BAO/$f\sigma_8$ + $S_8$ & w/ EFTofBOSS  & w/ EFTofBOSS + $S_8$ \\
\hline
{\it Planck} high-$l$ TTTEEE & 2347.7& 2347.3 & 2348.0 \\

{\it Planck} low-$l$ EE &397.2 & 396.9 & 397.0 \\

{\it Planck} low-$l$ TT & 23.1& 23.0  & 23.2 \\

{\it Planck} lensing & 8.9& 8.9 &  9.1 \\

Pantheon & 1027.2& 1027.1 & 1027.2 \\

Ext-BAO &6.1 & 6.2 &  6.2 \\

BOSS BAO/$f\sigma_8$  & 7.1& $-$ &  $-$ \\

EFTofBOSS & $-$& 117.8 & 118.3 \\

$S_8$ & 0.2& $-$ & 0.4 \\

\hline
total $\chi^2_{\rm min}$ &  3817.5 & 3927.0	&	3929.2	\\
\hline

\end{tabular}
\caption{$\chi^2$ of each data set for our `{\it Planck} + Pantheon + BOSS BAO/$f\sigma_8$ + Ext-BAO + $S_8$', `{\it Planck} + Pantheon + EFTofBOSS + Ext-BAO' and `{\it Planck} + Pantheon + EFTofBOSS + Ext-BAO + $S_8$' analyses for the DCDM $\to$ WDM+DR model. Since we rounded the $\chi^2$ of each experiment, the total $\chi^2$ is only equal to the sum of each $\chi^2$ at ${\cal O}(0.1)$ precision.}
\label{tab:chi2_total_WDM}
\end{table*}

\section{$\Lambda$CDM parameters of the DCDM $\to$ WDM+DR model}
\label{sec:app_WDM_LCDM}

\begin{figure*}
    \centering
\includegraphics[width=2\columnwidth]{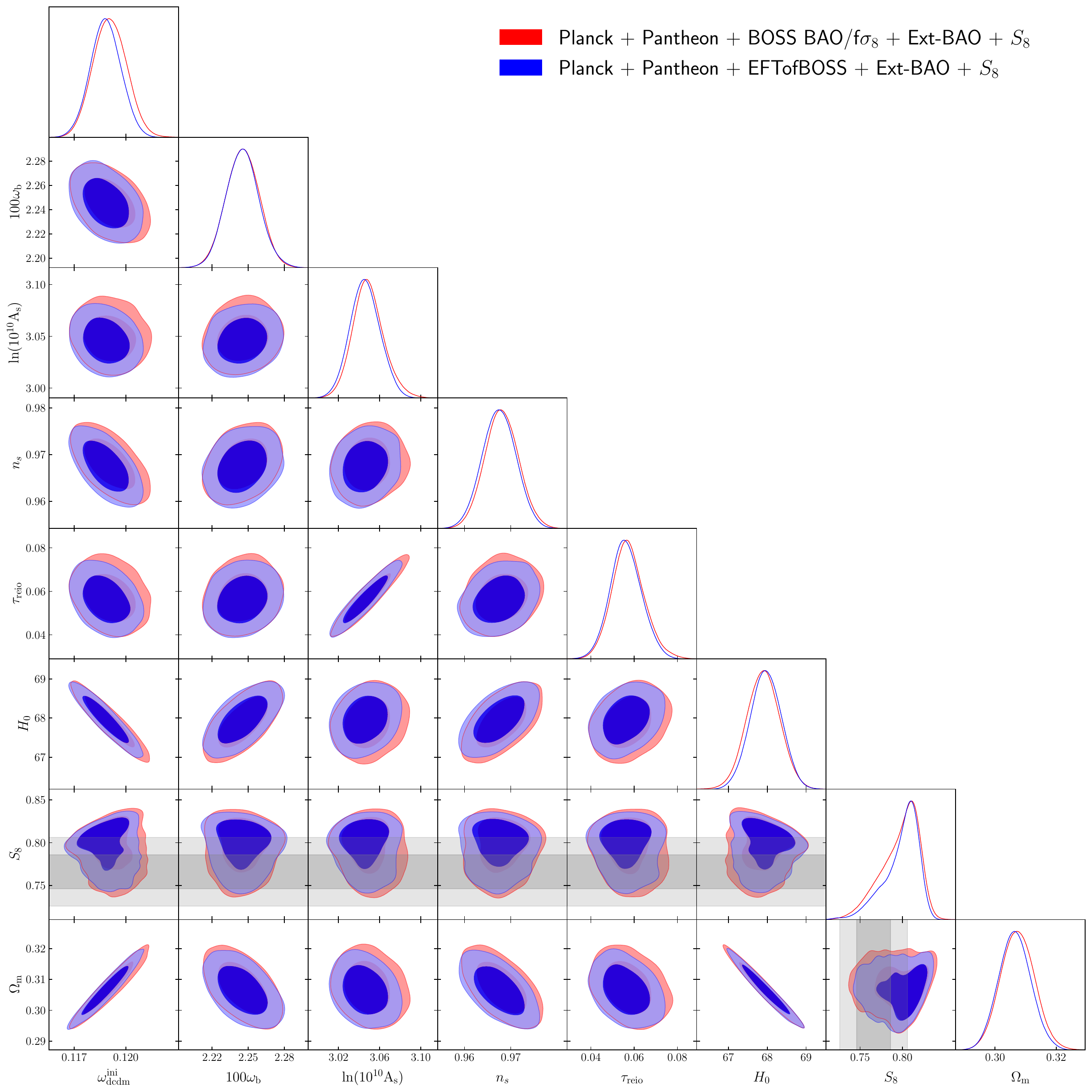}
\caption{2D posterior distributions of the DCDM $\to$ WDM+DR model with and without the EFTofBOSS data set for the $\Lambda$CDM parameters. We took into account the $S_8$ prior from KIDS-1000 for these two MCMC analyses. The gray shaded bands refer to the joint $S_8$ measurement from KIDS-1000 + BOSS + 2dFLens.  }
    \label{fig:appMCMCWDM1}
\end{figure*}

\begin{figure*}
    \centering
\includegraphics[width=2\columnwidth]{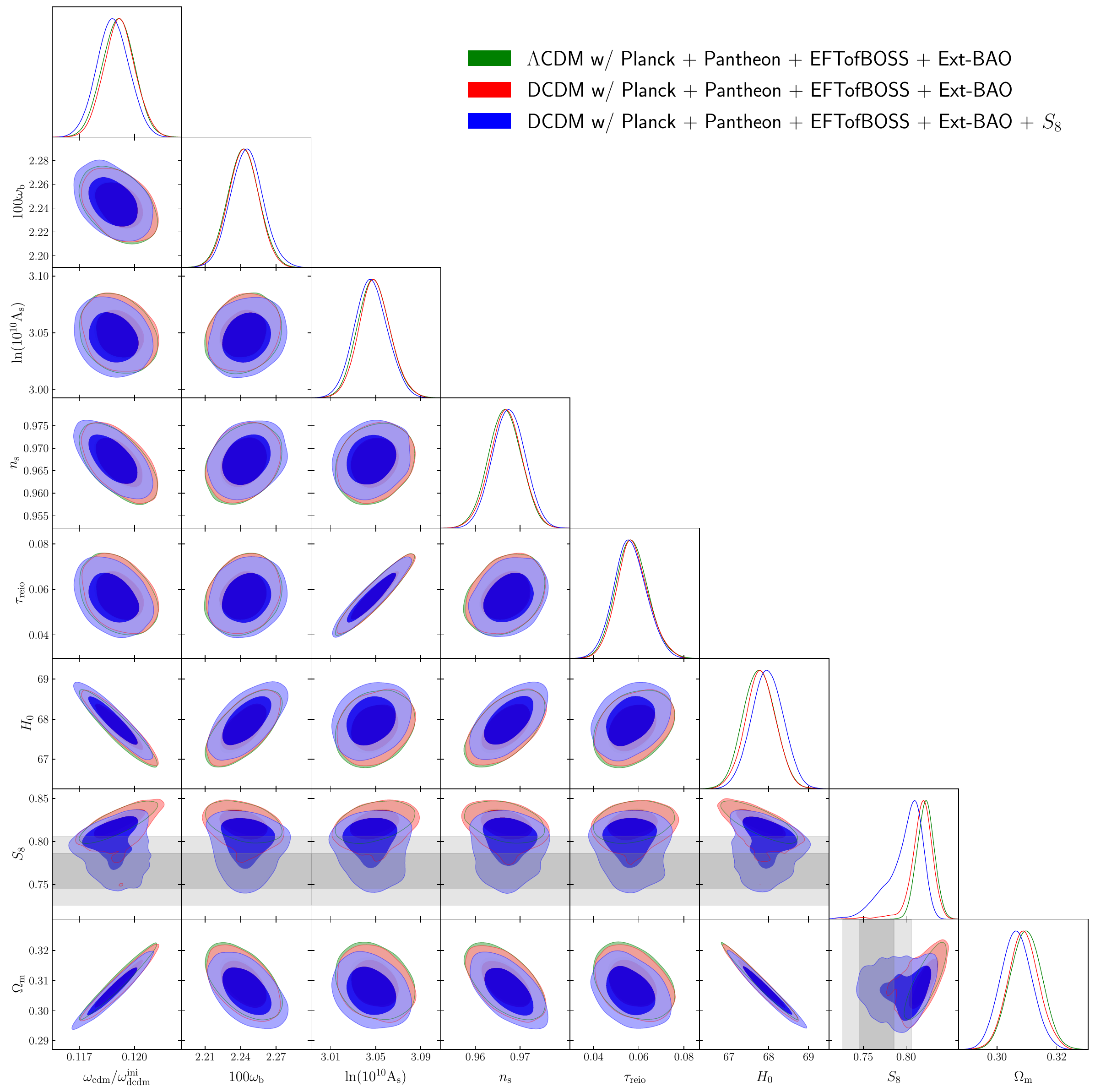}
\caption{2D posterior distributions of the DCDM $\to$ WDM+DR model with and without the $S_8$ prior from KiDS-1000. We also plot the 2D posterior distribution for the $\Lambda$CDM model without this $S_8$ prior. We took into account the EFTofBOSS data for these three MCMC analyses. The gray shaded bands refer to the joint $S_8$ measurement from KiDS-1000 + BOSS + 2dFLens.}
    \label{fig:appMCMCWDM2}
\end{figure*}

In this appendix, we compare in Fig. \ref{fig:appMCMCWDM1} the $\Lambda$CDM parameters of the DCDM $\to$ WDM+DR model obtained from the analyses with (blue) and without (red) the EFTofBOSS data, while in Fig. \ref{fig:appMCMCWDM2} we represent the $\Lambda$CDM parameters reconstructed from an analysis of {\it Planck}, Pantheon, Ext-BAO and EFTofBOSS data, with (blue) and without (red) the $S_8$ prior. 
We also show in this second figure the standard $\Lambda$CDM posteriors as a reference (green).
One can see that  the $\Lambda$CDM parameters are left largely unchanged in the DCDM $\to$ WDM+DR model. 
The decay into warm products only affect the growth of structure at late-times with little impact on parameters that could affect early time physics. 
This is essentially why cosmological data other than those measuring $S_8$ (and potentially the growth of structure at late-times) are unaffected by the DCDM, despite a lower $S_8$ value.

\newpage
\bibliography{dcdm}

\end{document}